\documentclass[12pt]{article}
\usepackage{a4wide}
\usepackage{graphicx}
\usepackage{amssymb}

\newcommand{\be}{\begin{equation}}
\newcommand{\ee}{\end{equation}}
\newcommand{\ba}{\begin{eqnarray}}
\newcommand{\ea}{\end{eqnarray}}

\newcommand{\vdir}{v\kern -5.75pt\raise 0.15ex\hbox {/}}
\newcommand{\Tr}[1]{\mathrm{Tr}\left[#1\right]}
\begin{document}

\begin{titlepage}
\begin{flushright}
LU TP 10-27\\
November 2010
\end{flushright}
\vfill
\begin{center}
{\Large\bf Vector Formfactors in Hard Pion\\[3mm]Chiral Perturbation Theory}
\vfill
{\bf Johan Bijnens and Ilaria Jemos}\\[0.3cm]
{Department of Astronomy and Theoretical Physics, Lund University,\\
S\"olvegatan 14A, SE 223-62 Lund, Sweden}
\end{center}
\vfill
\begin{abstract}
We use three-flavour hard pion Chiral Perturbation Theory (HPChPT)
in both the heavy meson and a relativistic formulation
to calculate the chiral logarithms $m^2\log{\left(m^2/\mu^2\right)}$
contributing to the formfactors of the $B_{(s)}\rightarrow \pi,K,\eta$ and
$D_{(s)}\rightarrow \pi,K,\eta$ transitions
at momentum transfer $q^2$ away from the endpoint 
$q^2_\mathrm{max}=(m_B-m_M)^2$. 
We compare our results with CLEO $D\to\pi$ and $D\to K$ data.
We also calculate the Isgur-Wise function of the
$B_{(s)}\rightarrow D_{(s)}$ semileptonic decay away from the endpoint and the
chiral logarithms for the pion and kaon electromagnetic formfactor.

In two-flavour HPChPT
we calculate the chiral logarithms
for the pion vector and the scalar formfactors
at $s\gg m_\pi^2$. This allows
us to test hard pion ChPT using the existing two-loop calculations for these
quantities.
\end{abstract}
\vfill
\end{titlepage}

\section{Introduction}
\label{introduction}

As a result of the rapid progress in computer technology, simulations of full
QCD on the lattice are becoming increasingly feasible and thus many results
are now available. To improve their precision it is important to acquire
control on all the sources of systematic errors involved. One of them is
due to the fact that most simulations are done
with meson masses larger then the physical ones. It is therefore
essential to perform a chiral extrapolation of the lattice data points to
achieve smaller meson masses.

In this respect Chiral
Perturbation Theory (ChPT) \cite{Weinberg0,GL1,GL2}, the effective field
theory of QCD at low energy, plays a key role. This theory can predict the
dependence on the light quark masses of the observables under study,
via a systematic expansion
in the masses and momenta of the light mesons
using both
the spontaneous and explicit breaking of chiral symmetry.
Unfortunately ChPT is often limited by its range of validity. There exist
several processes where it is applicable only in a small fraction of the
allowed range of energy, while the extrapolations formulas are needed
elsewhere. It is the case for example of the $K\rightarrow \pi$, $D\rightarrow
\pi,K,\eta$ and $B\rightarrow \pi,K,\eta$
transition formfactors in e.g. semileptonic decays. These processes are very
important for the determination of CKM matrix elements, obtained combining
knowledge on the amplitudes from experiments
\cite{:2008yi,Adam:2007pv,Hokuue:2006nr,Aubert:2006px}  and the formfactors
calculated on the lattice \cite{Gamiz:2008iv}. The matching between lattice and
experimental data is done when the momentum transfer squared
to the vector boson is
small, i.e. when a hard external pion arises and thus the power counting
scheme of ChPT breaks down.
However it is possible to exploit the chiral symmetry of QCD
even there and predict the dependence on the meson masses of
the formfactors using the arguments of hard pion Chiral Perturbation Theory
(HPChPT).

This was first studied in \cite{Flynn:2008tg} where it was
applied to the semileptonic decay  $K\rightarrow \pi$ using two-flavour ChPT.
 They
argued there that it is possible to calculate the corrections of the type
$m^2_\pi\log{m^2_\pi/\mu^2}$ even when the squared momentum transfer $q^2$ is
very small, i.e. when the outgoing pion is hard. Their arguments are based on
the fact that only the soft internal pions are responsible for the chiral
logarithms.  These ideas have then been generalized and applied to
$K\rightarrow\pi\pi$ \cite{Bijnens:2009yr} and to
$B\rightarrow\pi\ell\nu_\ell$ \cite {Bijnens:2010ws}, always in the framework
of two-flavour ChPT. In \cite{Bijnens:2009yr,Bijnens:2010ws} it was made clear
that
the underlying arguments correspond basically
to separate the hard-structure of a
Feynman diagram from the soft one and use this last one to calculate infrared
singularities. The arguments are essentially the same as those used for photon
infrared singularities.

In this paper we use two-flavour HPChPT for the vector and the scalar
formfactors of the pion $F^\pi_V(s)$ and $F^\pi_S(s)$
at $s\gg m_\pi^2$. Our main interest in this calculation is that it
allows a test of the arguments of HPChPT using the existing two-loop
results in standard two-flavour ChPT \cite{Bijnens:1998fm}.
For completeness we also quote the three-flavour HPChPT results for
the electromagnetic formfactor for pions and kaons.

The main new result of this work is the three-flavour HPChPT
calculation of the transition formfactors in vector transitions
of $B$ and $D$ to $\pi,K$ and $\eta$ and the Isgur-Wise function
in $B$ to $D$ transitions. In the latter case, the calculations did exist
and has been used but the validity of the formulas was not discussed.
Our results improve the comparison between the measured $D\to\pi$
and $D\to K$ formfactors \cite{:2008yi}. We also calculate the contributions
at the endpoint for all these transitions where the results for the transitions
to $\eta$ are new.

The paper is organized as follows. In Sect.~\ref{preliminaries}
we briefly review ChPT and heavy meson ChPT (HMChPT). We present
the relativistic theory that has been used to
have an extra check of the correctness of our results here as well.
At the end of this
section we also summarize the arguments why HPChPT works, although we refer
the reader for further details to Sect.~5 of \cite{Bijnens:2010ws}. 

In Sect.~\ref{piK} we present the results for pion and kaon formfactors
and test HPChPT using the existing two-flavour two-loop calculations
for the pion formfactors.
In Sect.~\ref{Btolight} we
define the formfactors of the heavy to light transitions
and present our results for them. We also show 
here the comparison with the experimental data from \cite{:2008yi}
on the $D\rightarrow\pi(K)$ transitions .
 The $B\to D$ transitions are defined
and our results for them given in Sect.~\ref{BtoD}.
In the appendix we provide some
results for the needed expansions of the loop integrals.

\section{Chiral Perturbation Theory}
\label{preliminaries}

\subsection{Standard Chiral Perturbation Theory}

In this subsection we briefly describe the formalism of ChPT
\cite{Weinberg0,GL1,GL2} for both two- and three-flavour ChPT.
The notation in the following is the same as in \cite{Bijnens:1999sh}.
The lowest order Lagrangian describing the strong interactions of the light
mesons is
\be
\label{pilagrangian}
\mathcal{L}_{\pi \pi}^{(2)}= \frac{F^2}{4}  
\left( \langle u_{\mu} u^{\mu} \rangle   + \langle \chi_{+}\rangle \right),
\ee
with
\ba
 u_{\mu} &=& i\{  u^{\dag}( \partial_{\mu} - i r_{\mu}   )u
 -u ( \partial_{\mu}   -i l_{\mu}    ) u^{\dag}   \}\,,
\nonumber\\
\chi_{\pm} &=& u^{\dag} \chi u^{\dag} \pm u \chi^{\dag} u\,,
\nonumber\\
u &=& \exp\left(\frac{i}{\sqrt{2}F} \phi \right)\,,
\nonumber\\
 \chi &=& 2B (s+ip)\,.\nonumber
\ea
$u$ parametrizes the oscillations around the vacuum in
$SU(n)_L\times SU(n)_R/SU(n)_V\sim SU(n)$ for $n=2,3$ the number
of light flavours. $\phi$ is thus a hermitian $n\times n$
matrix:
\ba
\phi =
 \left( \begin{array}{ccc}
\frac{1}{\sqrt{2}}\pi^0  + \frac{1}{\sqrt{6}} \eta &  \pi^+ & K^+ \\
 \pi^- & -\frac{1}{\sqrt{2}}\pi^0 + \frac{1}{\sqrt{6}} \eta  & K^0 \\
 K^- & \bar{K}^0 & -\frac{2}{\sqrt{6}} \eta  \\
 \end{array} \right)\,,\qquad
\phi =
 \left( \begin{array}{cc}
\frac{1}{\sqrt{2}}\pi^0  &  \pi^+  \\
 \pi^- & -\frac{1}{\sqrt{2}}\pi^0 \\
 \end{array} \right)\,.
\ea
The fields $s$, $p$, $l_\mu=v_\mu-a_\mu$ and
$r_\mu=v_\mu+a_\mu$ are the standard external scalar, pseudoscalar, left- and
right-handed vector fields introduced by Gasser and Leutwyler \cite{GL1,GL2}.
We will use the symbol $F$ throughout this paper but it should
be kept in mind that $F$ can be either the two-flavour constant
called $F$ in \cite{GL1} or the three-flavour one called $F_0$ in \cite{GL2}.

The field $u$ and $u_\mu$ transform under a chiral transformation $g_L\times g_R
\in SU(n)_L\times SU(n)_R$ as
\ba
\label{trasfrules}
u \longrightarrow g_R u h^\dagger = h u g_L^\dagger,\qquad
u_\mu\longrightarrow h u_\mu h^\dagger.
\ea
In (\ref{trasfrules}) $h$ depends on $u$, $g_L$ and $g_R$ and
is the so called compensator field.
The notation $\langle X\rangle$ stands for trace over up,down quark
indices for $n=2$ and up, down, strange for $n=3$.

Starting from this Lagrangian we can then build an effective field theory
by including loop diagrams and higher order Lagrangians. Introductions to
ChPT can be found in \cite{Pich,Scherer1}.

\subsection{Heavy meson Chiral Perturbation Theory}
\label{HMChPT}

In this subsection we briefly describe the formalism of HMChPT
\cite{Wise:1992hn,Burdman:1992gh,Goity:1992tp}.
Longer introductions can be found in the
lectures by Wise  \cite{Wise:1993wa} and the book \cite{Heavyquarkbook}.

The combination of Heavy Quark Effective Theory and
of ChPT provides us with a powerful formalism to study hadrons
containing a heavy quark. This combination is called HMChPT. It makes
use of both
spontaneously broken $SU(n)_L\times SU(n)_R$ chiral symmetry on the light
quarks, and spin-flavour symmetry on the heavy quarks. 
Thus HMChPT involves both a heavy
and a light scale. 
The first one is the heavy meson mass and rules an expansion in powers of its
inverse. The second is the light meson mass that lets us study
chiral symmetry breaking effects in a chiral-loop fashion as in standard
ChPT.

The sector of the Lagrangian involving only light-quarks
has already been discussed above. We now present the
heavy meson part of the HMChPT Lagrangian for the
three-flavour case \cite{Wise:1992hn,Burdman:1992gh,Goity:1992tp}.
Hereafter we
concentrate on the $B^{(*)}$ mesons, but the same equations hold for the 
$D^{(*)}$ mesons as well. In the limit $m_b\rightarrow\infty$,
the pseudoscalar, $B$, and the vector, $B^*$, mesons are degenerate.
All results in this paper are in the leading order in the heavy quark
expansion.
Thus in the following we neglect the mass splitting $\Delta=m_{B^*}-m_B$.
To implement the heavy quark symmetries it is convenient to assemble
them into a single field
\ba
\label{heavyfield}
H_a(v) &=& {1 +  \vdir \over 2} \left[ B^{*}_{a\mu} (v)\gamma^\mu - B_a
  (v)\gamma_5\right],
\ea
where $v$ is the fixed four-velocity of the heavy meson, $a$ is a flavour
index corresponding to the light quark in the heavy meson. Therefore
$B_1 = B^+$, $B_2=B^0$, $B_3=B_s$, while $D_1 = \overline{D^0}$,
$D_2=D^-$, $D_3=\overline{D_s}$
and similarly for the vector mesons $B^*_\mu$ and $D^*_\mu$.
In (\ref{heavyfield}) the operator
$(1+\vdir)/2$ projects out the particle component of the heavy meson only.
The conjugate field is defined as 
$ \overline H_a(v) \equiv \gamma_0 H_a^\dagger (v) \gamma_0$.
We assume the field
 $H_a(v)$ to
 transform under the chiral transformation $g_L\times g_R
\in SU(n)_L\times SU(n)_R$ as
\be
H_a(v) \longrightarrow h_{ab}H_b(v)\,,
\ee
so we introduce the covariant derivative 
\ba
D^\mu_{ab}H_b(v) = \delta_{ab} \partial^\mu H_b(v)  +\Gamma^\mu_{ab}H_b(v),
\ea
where 
$\Gamma^\mu_{ab} =
\frac{1}{2}\left[u^\dagger\left(\partial_\mu-ir_\mu\right)u+u\left(\partial_\mu-il_\mu\right)u^\dagger\right]_{ab}$, 
 and the indices $a$, $b$ run over the
light quark flavours. 
Finally, the 
Lagrangian for the heavy-light mesons in the static heavy quark limit reads 
\ba 
\label{heavylag}
\mathcal{L}_{\rm heavy}=-i\,\Tr{\overline{H}_a  v \cdot D_{ab}H_b}+g\,
\Tr{\overline{H}_a u_{ab}^\mu H_b \gamma_\mu \gamma_5},
\ea
where $g$ is the coupling of the heavy meson doublet to the Goldstone boson
and the traces, $\mathrm{Tr}$, are over spin indices,
the $\gamma$-matrix indices.
The Lagrangian (\ref{heavylag}) satisfies chiral symmetry and heavy
quark spin flavour
symmetry. We neglect in the following the mass differences for the
heavy mesons containing the same heavy quark.

\subsection{Relativistic theory}
\label{relth}

When the momentum transfer to the light degrees of freedom is not
small as in HPChPT,
very off-shell heavy mesons may appear in the loops.
Different
treatments of the off-shell behaviour modify the loop-functions.
Thus in principle it might change
the non-analyticities in the light meson masses.
If this were the case, the arguments
summarized in Sect.~\ref{HPChPT} would be wrong. In fact,
provided that the two formalisms are both
sufficiently complete, the soft singularities must be the same, since
they are arising in the same way. This is the reason why both
here and in \cite{Bijnens:2010ws} we are calculating not only using HMChPT
but also in a relativistic formulation as a check on the arguments.

We use a relativistic
Lagrangian that respects the
spin-flavour symmetries of HMChPT. It is essentially the same Lagrangian
introduced in \cite{Bijnens:2010ws}, but now in the three-flavour case. The $B_a$
and $B^{*}_{a\mu}$ fields are in the relativistic form and we treat
them as column-vectors in the light-flavour index $a$.
\ba
\label{rellagkin}
\mathcal{L}_{\rm kin}&=&
\nabla^\mu B^\dagger \nabla_\mu B -m_B^2 B^\dagger B
-\frac{1}{2}B^{*\dagger}_{\mu \nu}B^{*\mu \nu}
+m_B^2 B^{*\dagger}_{\mu}B^{*\mu},\\
\mathcal{L}_{\rm int} &=&g M_0\left(B^\dagger u^\mu
B^{*}_\mu+B^{*\dagger}_\mu u^\mu
B\right)\nonumber\\
&+&\frac{g}{2}\epsilon^{\mu\nu\alpha\beta}\left(-B^{*\dagger}_\mu
u_\alpha\nabla_\mu B^*_\beta+\nabla_\mu B^{*\dagger}_\nu
u_\alpha B^*_\beta\right),\label{rellagint}
\ea
with
$
B^*_{\mu \nu}=\nabla_\mu B^*_\nu-\nabla_\nu B^*_\mu$, and
$\nabla_\mu =
\partial_\mu+\Gamma_\mu$.
The constant $g$ of (\ref{rellagint}) is the same in
(\ref{heavylag}), $M_0$ is the mass of the $B$ meson in the chiral limit.
The fields $B$
and $B^*$ transform under chiral transformations as $B\to hB$.
The two terms of $\mathcal{L}_{\rm int}$ in (\ref{rellagint}) contain
the vertices $B B^{*} M$ and 
$B^{*}B^{*}M$ for $M=\pi,K,\eta$.

{}From $\mathcal{L}_{\rm kin}$ in (\ref{rellagkin}) we find the
propagators of the $B$ and $B^{*}$ meson respectively:
\be
\frac{i}{p^2-m_B^2}, \qquad \frac {-i\left(g_{\mu \nu}-\frac{p_\mu p_\nu}{m^2_B}\right)}{p^2-m^2_B}.
\ee
This is to be contrasted with the propagator $1/v\cdot p$ in the HMChPT
showing the different off-shell behavior.

\subsection{Hard pion Chiral Perturbation Theory}
\label{HPChPT}

In general,
the use of ChPT and HMChPT is valid as long as the interacting light mesons
are soft,
i.e. if they have momenta much smaller than the scale of spontaneous chiral
symmetry breaking ($\Lambda_{\rm ChSB}\simeq 1~{\rm  GeV}$). Only in this
regime is the power counting of ChPT well defined.

On the other hand the
arguments presented in great detail in Sect.~5 of \cite{Bijnens:2010ws} show
that the predictions of the soft singularities in the light meson masses
appearing
in the final amplitudes are reliable even outside the range of applicability
of HMChPT. Hereafter we present a short summary of these arguments, but for a
comprehensive discussion we refer the reader
to Sect.~5 of \cite{Bijnens:2010ws}.

The underlying idea is that in a loop diagram, the internal soft light 
mesons are the source of the infrared non-analyticities arising,
even if hard, i.e. large momentum, light mesons are present.
Since the soft lines do not see the hard or short-distance structure of
the diagram, we can separate them from the rest of the process. 
We should thus be able to describe the hard part of any
diagram by an effective Lagrangian which must include
the most general terms  consistent with all the symmetries. The
coefficients of this Lagrangian depend on the hard kinematical quantities and
can even be complex. This Lagrangian must be sufficiently complete to describe
the neighbourhood of the underlying hard process.

Extra caution must be taken to build up the Lagrangian describing the hard
part. As a matter of fact we can not neglect operators with an arbitrary
numbers of derivatives since the momenta into play can be large. However it
turns out that matrix elements of operators with higher number of
derivatives are all proportional to the lowest order ones up to terms
of higher order, i.e. the coefficients of the leading non-analyticities
are calculable in terms of the lowest order Lagrangians.

We expect that a full power counting can be formulated along the
lines of SCET \cite{SCET} but the leading prediction can be obtained
in the simpler fashion done here.

\section{Pion and kaon formfactors}
\label{piK}

\subsection{Electromagnetic formfactors in three-flavour HPChPT}

The vector (electromagnetic) formfactors of the charged
pion and kaon are defined as
\ba
\label{FpiV:def}
\left< \pi(K)^+(p_2)\left|j^{\rm elm}_\mu \right |\pi(K)^+(p_1) \right>&=&(p_2+p_1)_\mu F^{\pi(K)}_V(s),
\ea
with $s=(p_1-p_2)^2$ and
 $j^{\rm elm}_\mu=\frac{2}{3}\bar{u}\gamma_\mu
u-\frac{1}{3}\bar{d}\gamma_\mu d-\frac{1}{3}\bar{s}\gamma_\mu s$
is the electromagnetic
current.
The arguments of HPChPT can be used here as well and we get from
the relevant one-loop diagrams and wave function renormalization that
\ba
\label{FVthree}
F^\pi_V(s) &=& 
F^{\pi\chi}_V(s)\left(1+\frac{1}{F^2}\overline A(m_\pi^2)+\frac{1}{2F^2}\overline A(m_K^2)+\mathcal{O}(m_L^2)\right)\,,
\nonumber\\
F^K_V(s) &=& 
F^{K\chi}_V(s)\left(1+\frac{1}{2F^2}\overline A(m_\pi^2)+\frac{1}{F^2}\overline A(m_K^2)+\mathcal{O}(m_L^2)\right)\,.
\ea
The superscript $\chi$ here means in the limit $m_u=m_d=m_s=0$.
In the remainder we will usually drop the $\mathcal{O}(m_L^2)$ part
but all results should be interpreted as up to analytic terms
in the light meson masses squared. The loop integral $\overline{A}(m^2)$
is defined in the appendix. The result (\ref{FVthree}) can be calculated
directly or by expanding the known ChPT result \cite{GL3,BC} for
$s\gg m_L^2$.

\subsection{Vector and scalar pion formfactors in two-flavour HPChPT}
\label{twoloops}

It is important to test the arguments behind HPChPT as much as possible.
We can do a nontrivial test by looking at the two-flavour case for
the pion vector and scalar formfactors. The vector form factor is defined
in (\ref{FpiV:def}) and the scalar formfactor is defined by
\ba
\label{FpiS:def}
\left< \pi^0(p_2)\left|\bar{u}u+\bar{d}d\right|\pi^0(p_1)
\right>&=&f^{\pi}_S(0)F^{\pi}_S(s)\,.
\ea
We have factored out here as is customary \cite{GL1,GL3,Bijnens:1998fm} the
value at $s=0$. From the general discussion we again expect that
the leading non-analytic correction should be in both cases of the
form
\ba
\label{genform}
f(s)=C(s)\times\left(1+\alpha \frac{m^2}{F^2}\log{\frac{m^2}{\mu^2}}
+ \mathcal{O}(m^2)\right)\,.
\ea
In principle $\alpha$ could depend on $s$ but it is calculable.
$C(s)$ is a free parameter in HPChPT and can even be complex.

Calculating the formfactors
from wave-function renormalization and the needed one-loop
diagrams we obtain
\ba
\label{FS:oneloop}
F^\pi_V(s) &=& 
F^{\pi\chi}_V(s)\left(1+\frac{1}{F^2}\overline A(m_\pi^2)\right)\,,
\nonumber\\
F^\pi_S(s) &=& 
F^{\pi\chi}_S(s)\left(1+\frac{5}{2F^2}\overline A(m_\pi^2)\right)\,.
\ea
Here $\chi$ means in the limit $m_u=m_d=0$.
This agrees with the large $s$ expansion of the one-loop result of \cite{GL1}.

In normal ChPT these formfactors are known fully analytically up till
two-loop order \cite{Bijnens:1998fm}.
We can now choose a value of $m_\pi^2$ and $s$ such that $s\gg m_\pi^2$
but with both $s$ and $m_\pi^2$ in the regime of validity of standard HPChPT.
The expansion for $s\gg m_\pi^2$ can be done and the result should be of the
form (\ref{FS:oneloop}) where the form of $F^{\pi\chi}_V(s)$, $F^{\pi\chi}_S(s)$
follows from the one-loop calculation in the limit $m_\pi^2=0$.
This gives
\ba
\label{oneloopzero}
F^{\pi\chi}_V(s) &=& 1+\frac{s}{16\pi^2F^2}
\left(\frac{5}{18}-16\pi^2 l_6^r+\frac{i\pi}{6}-\frac{1}{6}\ln\frac{s}{\mu^2}
 \right)\,,
\nonumber\\
F^{\pi\chi}_S(s) &=& 1+\frac{s}{16\pi^2F^2}
\left(1+16\pi^2 l_4^r+i\pi-\ln\frac{s}{\mu^2}
 \right)\,.
\ea

Let us see what happens when the full two-loop results are taken into account. 
Our arguments still hold as long as we are working at the desired order
i.e. $\mathcal{O}(m^2_\pi)$. On the other hand now different kind of terms
arise. Some of them are suppressed by $m^4_\pi$ with or without
logarithms and so can be neglected. The ones like $s^2$ or $s^2\log{s^2/\mu^2}$
and without $\log(m^2_\pi/\mu^2)$ cannot be predicted by HPChPT and are
absorbed in the unknown part of the coefficient $C(s)$ of (\ref{genform}).
Terms like
$s^2\log^2{m^2_\pi/\mu^2}$ or $s^2\log{m^2_\pi/\mu^2}$
can also arise. Those
not only would be large in our limit, but even divergent when
$m_\pi\rightarrow 0$, therefore they must cancel. 
Terms like $s m^2_\pi \log^2{m^2_\pi/\mu^2}$ are predicted by HPChPT
not to occur. Finally
there are terms as $s
m^2_\pi \log(m^2_\pi/\mu^2)$ and $s
m^2_\pi \log(m^2_\pi/\mu^2) \log(s/\mu^2)$ which are of special interest. The
coefficients of these are predicted by HPChPT. They are given by
(\ref{oneloopzero}) and (\ref{FS:oneloop}).
Performing the expansion of the full two-loop result for $s\gg m_\pi^2$
we indeed find that the result is of the required form with the chiral limit
value given exactly by (\ref{oneloopzero}). This is a rather nontrivial check
on HPChPT.

\section{$B\rightarrow M$ and $D\rightarrow M$ transitions}
\label{Btolight}

\subsection{Definition of formfactors}
\label{Bl3:formalism}

In this section we review the formalism for the transitions of a
$B$ or a $D$ meson into a light pseudoscalar meson ($\pi$, $K$, $\eta$).
We restrict ourselves to
the case of a $B$ meson, but the same definitions hold also for the
$D$-decay. All the following discussion can be found also in
\cite{Bijnens:2010ws} for the two-flavour case.
We report it here for the sake of completeness.
The hadronic current for pseudoscalar to pseudoscalar vector transitions
($P_i(\bar{q_i},q)\rightarrow P_f(\bar{q_f},q)$) has the
structure
\ba
\label{QCDformfact}
\left< P_f(p_{f}) \left|\overline{q}_i \gamma_\mu q_f\right|P_i(p_i)\right>
&=&(p_i+p_f)_\mu f_+(q^2)+ (p_i-p_f)_\mu f_-(q^2)\\
&=&\left[(p_i+p_f)_\mu -q_\mu\frac{(m^2_i-m^2_f)}{q^2}\right]f_+(q^2)
+q_\mu\frac{(m^2_i-m^2_f)}{q^2}f_0(q^2),\nonumber
\ea
where $q^\mu$ is the momentum transfer $q^\mu=p^\mu_i-p^\mu_f$. 
In our case $P_f$ is a light pseudoscalar meson, $P_i$ is a $B$ meson and
$q_i=b$. 
For example, to find the $B^0 \rightarrow \pi^+$
formfactors we need then to
evaluate the hadron matrix elements of the quark bilinear $\overline{b}
\gamma_\mu q$, where $q=u$.

Parity invariance, heavy quark and chiral symmetry dictate
that the matching of QCD
bilinears onto operators of HMChPT take the form \cite{Falk:1993fr,Wise:1993wa},
\be
\label{matching}
\bar{b}\gamma^\mu q_a\propto c\left\{\Tr{\gamma^\mu
  \left(u^\dagger_{ab}+u_{ab}\right)H_b(v)}
+\Tr{\gamma_5\gamma^\mu \left(u^\dagger_{ab}-u_{ab}\right)H_b(v)}\right\}.
\ee
If no hard pions appear in the final state we can use the
definition of the decay constant
\be
\left< 0 \left|\overline{b} \gamma_\mu \gamma_5 q\right|{B}(p_B)\right>
= i F_B p^\mu_B 
\ee
and obtain $c=\frac{1}{2}F_B \sqrt{m_B}$. This latter result
does not hold for momenta away from $q^2_{max}$ in which case $c$ is just
an effective coupling depending on $q^2$.

In HMChPT the definitions of the
formfactors are chosen such that those are independent of the heavy meson
mass. So for example for the $B\rightarrow M$ transition
\ba\label{HMChPTformfact}
\left< M(p_M) \left|\overline{b} \gamma_\mu q\right|B(v)\right>_{\rm HMChPT}
=\left[p_{M\mu}-\left(v\cdot p_M\right)v_\mu\right] f_p(v\cdot p_M)
+ v_\mu f_v(v\cdot p_M).
\ea
In (\ref{HMChPTformfact}) $v\cdot p_M$ is the energy of the light meson
in the heavy meson rest frame
\be
v\cdot p_M=\frac{m^2_B+m^2_{M}-q^2}{2m_B}.
\ee
The formfactors defined in~(\ref{QCDformfact}) and in~(\ref{HMChPTformfact})
are related by matching the relativistic and the HMChPT hadronic current:
\ba\label{relationsff1}
\sqrt{m_B}f_p(v\cdot p_M)&=&
f_+(q^2)+\frac{m^2_B-m^2_M}{q^2}f_+(q^2)
-\frac{m^2_B-m^2_M}{q^2}f_0(q^2)\nonumber\\
&=&f_+(q^2)-f_-(q^2),\\
\label{relationsff2}
\sqrt{m_B}\left(f_v(v\cdot p_M)-f_p(v\cdot p_M)v\cdot p_M\right)&=&
m_B\left( \frac{q^2-m^2_B+m^2_M}{q^2}f_+(q^2)
+\frac{m^2_B-m^2_M}{q^2}f_0(q^2)\right)\nonumber\\
&=&m_B\left(f_+(q^2)+f_-(q^2)\right).
\ea
The $\sqrt{m_B}$ factors  in (\ref{relationsff1}) and (\ref{relationsff2}) are due to the different
normalizations of states used in the two formalisms.
At $q^2\approx q^2_{\rm max}$, neglecting terms suppressed by powers of
$m_M$ and of $1/m_B$, (\ref{relationsff1}) and (\ref{relationsff2}) become
\be
\label{relationsffqmax}
f_0(q^2)=\frac{1}{\sqrt{m_B}} f_v(v\cdot p_M), \qquad
f_+(q^2)=\frac{\sqrt{m_B}}{2}f_p(v\cdot p_M).
\ee
We remark that the relations in (\ref{relationsffqmax})
are  valid only
when $q^2\approx q^2_{\rm max}$
contrary to what was said in the original version\footnote{Notice that this
does not invalidate the results of \cite{Bijnens:2010ws}. Indeed all the
formfactors involved have the same chiral logarithms, thus the tree-level part
can still be factorized out, as shown in Sect.~\ref{Bl3:ChLogs}} of
\cite{Bijnens:2010ws}. At general $q^2$ away from $q^2_{\rm  max}$ we must
use (\ref{relationsff1}) and (\ref{relationsff2}).

A matching similar to (\ref{matching}) has to be done also for
the relativistic theory described in
Sect.~\ref{relth}. We identify four possible operators\footnote{The last one
is higher order but we included it since it has a different type of contraction
of the Lorentz indices and as an explicit check on the arguments of HPChPT \cite{Bijnens:2010ws}.}
\ba
\label{relleftcurr}
J^L_\mu=\frac{1}{2} E_1 t  u^\dagger \nabla_\mu B +  \frac{i}{2} E_2 
t u^\dagger u_\mu B +  \frac{i}{2} E_3 t u^\dagger B^{*}_\mu+ \frac{1}{2}
E_4 t  u^\dagger \left(\nabla_\nu u_\mu \right) B^{* \nu},
\ea
where $E_1$,$\dots$, $E_4$, are effective couplings. $t$ is a constant
spurion vector transforming as $t\rightarrow t g_L^\dagger$,
so that $J^L_\mu$ is invariant under $SU(3)_L$ transformations.
The heavy quark symmetry implies $m_BE_1=E_3$.
Analogously we can introduce a right-handed
$J^R_\mu$ current and thus an axial-vector $J^5_\mu=J^R_\mu-J^L_\mu$ and a
vector $J^V_\mu=J^R_\mu+J^L_\mu$ current. They are used respectively to
evaluate the amplitudes of $B\rightarrow \ell \nu_\ell$ and the
$B\rightarrow M$ formfactors as defined in (\ref{QCDformfact}).
We leave the discussion for the latter in Sects.~\ref{Bl3:ChLogs}
and \ref{Bl3:endpoint}, while we quote
here the expression of the $B(B_s)$ decay constants that can be found
evaluating the $B(B_s) \rightarrow$ vacuum matrix
element at one loop:
\ba
\label{Bdecayconst}
F_B&=&E_1 \left\{1+
       \frac{1}{F^2}
       \left[\left(\frac{3}{8}+\frac{9}{8}g^2\right)\overline{A}(m^2_\pi)+
       \left(\frac{1}{4}+\frac{3}{4}g^2\right)\overline{A}(m^2_K)+
       \left(\frac{1}{24}+\frac{1}{8}g^2\right)\overline{A}(m^2_\eta)\right]
\right\},\nonumber \\
\\
\label{Bsdecayconst}
F_{B_s}&=&E_1 \left\{1+
       \frac{1}{F^2}
       \left[\left(\frac{1}{2}+\frac{3}{2}g^2\right)\overline{A}(m^2_K)+
       \left(\frac{1}{6}+\frac{1}{2}g^2\right)\overline{A}(m^2_\eta)\right]
\right\}.
\ea
$\overline{A}(m^2)$ is defined in (\ref{A}) in the appendix.
Here we only quote
the non-analytic dependence on the light quark masses for the one-loop part.
The results
 (\ref{Bdecayconst}) and (\ref{Bsdecayconst}) agree with those obtained with
HMChPT \cite{Goity:1992tp}. We
see that $E_1$ plays the role of $F_{H}$ in \cite{Goity:1992tp}
and that the relativistic theory
predicts the same coefficient of the chiral logarithm in $\overline{A}(m^2)$
as expected.

\subsection{The chiral logarithms away from the endpoint}
\label{Bl3:ChLogs}

\begin{figure}
\begin{center}
\includegraphics{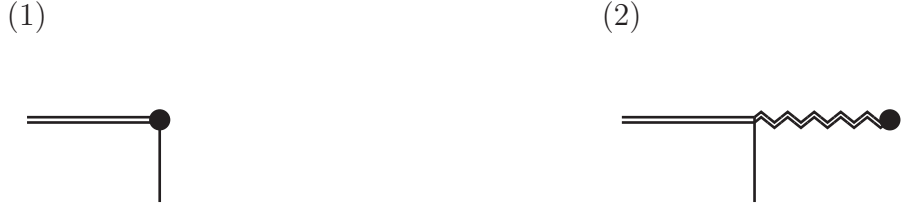}
\end{center}
\caption{The tree-level diagrams contributing to the amplitude. A double line
  corresponds to a $B$, a zigzag line to a $B^*$, a single line to a light
  meson, i.e. $\pi$, $K$ or $\eta$. A black circle represents the insertion
  of a vector current.}
\label{fig:treelevel}
\end{figure}

In this section we present results for the formfactors of the
vector transitions
$B\rightarrow\pi$, $B\rightarrow K$, $B\rightarrow
\eta$, $B_s\rightarrow K$ and $B_s\rightarrow\eta$
calculated using three-flavour HPChPT. The results for
the $B\rightarrow\pi$ transition in two-flavour ChPT can be found in
\cite{Bijnens:2010ws}. We quote only the relevant
terms, i.e. the leading ones which contain free parameters
and the predicted chiral logarithms
up to $\mathcal{O}(m^{2}_{M})$.
The tree-level diagrams contributing to the amplitude are shown in
Fig.~\ref{fig:treelevel}.
The formfactors at tree level read for HMChPT
\be
\label{HMChPTtree}
f^{\rm Tree}_p(v\cdot p_M)=C_{B\rightarrow M}
\frac{\alpha}{F}\frac{g}{v\cdot p_M+\Delta},
\qquad
f^{\rm Tree}_v(v\cdot p_M)=C_{B\rightarrow M}\frac{\alpha}{F},
\ee
where $C_{B\rightarrow M}$ is a constant that changes depending on the meson
transition and takes the values
\be
C_{B\rightarrow M}=\left\{ \begin{array}{ll}
1 & B^-\rightarrow\pi^0 \nonumber \\
\sqrt{2} & \bar{B^0}\rightarrow\pi^+ \nonumber \\
\sqrt{2} & B\rightarrow K \nonumber \\
\frac{1}{\sqrt{3}} & B\rightarrow \eta \nonumber\\
\sqrt{2} & B_s\rightarrow K \nonumber \\
-\frac{2}{\sqrt{3}} & B_s\rightarrow\eta. \nonumber \\
\end{array}
\right.
\ee
In (\ref{HMChPTtree}) $\alpha$ is a constant that takes the value
$\sqrt{m_B/2} F_B$ at $q^2_{\rm max}$. We also obtain $c=\alpha/\sqrt{2}$.
Near $q^2_{max}=(m_B-m_M)^2$ the results remain obviously the same, but the
propagator in the first equation of (\ref{HMChPTtree}) becomes $1/m_M$.
In the case of the relativistic theory of Sect.~\ref{relth}, we distinguish
the formfactors for the two $q^2$ ranges.
At $q^2$ away from $q^2_{\rm max}$
\ba\label{relthtreegen}
f^{\rm Tree}_+(q^2)\!\!\!\!&=&\!\!\!\!
C_{B\rightarrow M}\left\{-\frac{1}{4}\frac{E_3}{F}\frac{m_B}{q^2-m^2_B}g
+\frac{1}{8}\frac{E_1}{F}-\frac{1}{4}\frac{E_2}{F}\right\},
\nonumber\\
f^{\rm Tree}_0(q^2)\!\!\!\!&=&\!\!\!\!C_{B\rightarrow M}\left\{
\frac{1}{8}\frac{E_1}{F}\left(1+\frac{q^2}{m^2_B-m^2_M}\right)
-\frac{1}{4}\left(\frac{E_2}{F}+\frac{E_3}{F}\frac{m_B}{q^2-m^2_B}g
\right)
\left(1-\frac{q^2}{m^2_B-m^2_M}\right)\!\!\right\}.\nonumber\\
\ea
At $q^2\approx q^2_{\rm max}$ (\ref{relthtreegen}) simplifies to
\be\label{relthtreeqmax}
f^{\rm Tree}_+(q^2)_{q^2\approx q^2_{\rm max}}=
C_{B\rightarrow M}\frac{1}{4}\frac{E_3}{F}\frac{1}{2m_M}g,
\qquad
f^{\rm Tree}_0(q^2)_{q^2\approx q^2_{\rm max}}=
C_{B\rightarrow M}\frac{1}{4}\frac{E_1}{F}.
\ee
We stress once more that the relation of $E_1$ and $E_3$ to $F_B$ holds only
when $q^2\approx q^2_{max}$. As the momentum transfer is out of
this range the coupling constant are different at the different
values of $q^2$ and can even be complex.

\begin{figure}[htb]
\begin{center}
\includegraphics{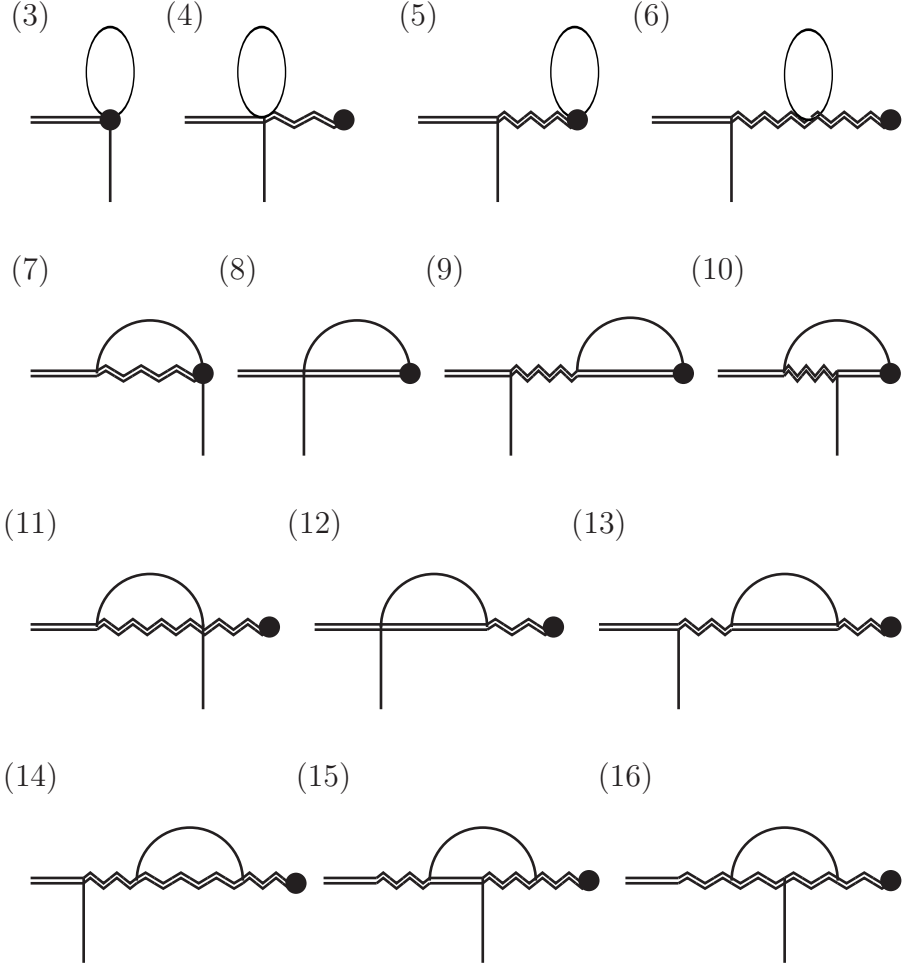}
\end{center}
\caption{The one-loop diagrams contributing to the amplitude. Vertices and
  lines as in Fig.~\ref{fig:treelevel}}
\label{fig:oneloop}
\end{figure}

At one-loop we need to include the contributions of the wavefunction
renormalization $Z_{\pi}$, $Z_{K}$, $Z_\eta$, $Z_B$ and $Z_{B_s}$. 
They are the same for HMChPT and the
relativistic theory and read:
\ba
Z_{\pi}&=&1-\frac{2}{3}\frac{\overline{A}(m^2_\pi)}{F^2}
-\frac{1}{3}\frac{\overline{A}(m^2_K)}{F^2},
\nonumber\\
Z_{K}&=&1-\frac{1}{4}\frac{\overline{A}(m^2_\pi)}{F^2}
-\frac{1}{2}\frac{\overline{A}(m^2_K)}{F^2}
-\frac{1}{4}\frac{\overline{A}(m^2_\eta)}{F^2},
\nonumber\\
Z_{\eta}&=&1-\frac{\overline{A}(m^2_K)}{F^2},
\nonumber\\
Z_{B}&=&1+\frac{9}{4}g^2\frac{\overline{A}(m^2_\pi)}{F^2}
+\frac{3}{2}g^2\frac{\overline{A}(m^2_K)}{F^2}
+\frac{3}{12}g^2\frac{\overline{A}(m^2_\eta)}{F^2},
\nonumber\\
Z_{B_s}&=&1+3g^2\frac{\overline{A}(m^2_K)}{F^2}
+g^2\frac{\overline{A}(m^2_\eta)}{F^2}.
\ea
The one-loop corrections to the vector current $J^V_\mu$ are shown in
Fig.~\ref{fig:oneloop}.

To find the results in HMChPT we
expanded the one-loop calculation of \cite{Becirevic:2003ad}
at $v\cdot p_M\rightarrow m_B$, $m^2_M\rightarrow0$. Note however that
their results are only valid near the endpoint. The arguments of HPChPT allow
us to use their results also away from the endpoint.

In the relativistic theory we first calculated
the formfactors and then we expanded the loop integrals
for $m^2_M \ll m^2_B, (m_B^2-q^2)$. These latter expansions are given in
App.~\ref{appendix}. Notice that we keep terms of the kind $m/M$ in the
expansion of the $\bar{C}$
and $\bar{B}$ functions (\ref{Cexp2}), (\ref{Bexp})
that had not been included explicitly in
\cite{Bijnens:2010ws}.  Those terms could cause corrections like $mM/F^2$ in
the final results, that would violate the heavy quark limit $M\to\infty$.
We verified
that all these corrections do cancel. 
To achieve the final results, we sum up all
the contributions coming from the several diagrams and include the
wavefunction renormalization pieces. This corresponds
to sum $(1/2) Z_M$ for $M=\pi,K,\eta$ and
 $(1/2) Z_B$ or $(1/2) Z_{B_s}$,
according to the external legs of the process under study,
multiplied by the tree-level part of the formfactors. 
We find for the different transitions that the two formfactors
always have the same chiral logarithms. We can write the results in the
form
\be
\label{Bfpv}
f_{v/p}(v\cdot p_M) =
f^{\rm Tree}_{v/p}(v\cdot p_M)
F_{B\to M}
\ee
The chiral logarithms are in $F_{B\to M}$ and read for the different transitions
\ba
\label{Bp:finalresult}
F_{B\to\pi}&=&
1+\left(\frac{3}{8}
    +\frac{9}{8}g^2\right)\frac{\overline{A}(m^2_\pi)}{F^2}
+\left(\frac{1}{4}+\frac{3}{4}g^2\right)\frac{\overline{A}(m^2_K)}{F^2}
+\left(\frac{1}{24}+\frac{1}{8}g^2\right)\frac{\overline{A}(m^2_\eta)}{F^2},
\nonumber\\
\label{BK:finalresult}
F_{B\to K} &=&
1+\frac{9}{8}g^2\frac{\overline{A}(m^2_\pi)}{F^2}
+\left(\frac{1}{2}+\frac{3}{4}g^2\right)\frac{\overline{A}(m^2_K)}{F^2}
+\left(\frac{1}{6}+\frac{1}{8}g^2\right)\frac{\overline{A}(m^2_\eta)}{F^2}\,,
\nonumber\\
\label{Be:finalresult}
F_{B\to\eta} &=&
1+\left(\frac{3}{8}+\frac{9}{8}g^2\right)\frac{\overline{A}(m^2_\pi)}{F^2}
+\left(\frac{1}{4}+\frac{3}{4}g^2\right)\frac{\overline{A}(m^2_K)}{F^2}
+\left(\frac{1}{24}+\frac{1}{8}g^2\right)\frac{\overline{A}(m^2_\eta)}{F^2}\,,
\nonumber\\
\label{BsK:finalresult}
F_{B_s\to K} &=&
1+\frac{3}{8}\frac{\overline{A}(m^2_\pi)}{F^2}
+\left(\frac{1}{4}+\frac{3}{2}g^2\right)\frac{\overline{A}(m^2_K)}{F^2}
+\left(\frac{1}{24}+\frac{1}{2}g^2\right)\frac{\overline{A}(m^2_\eta)}{F^2}\,,
\nonumber\\
\label{Bse:finalresult}
F_{B_s\to\eta}&=&
1+\left(\frac{1}{2}+\frac{3}{2}g^2\right)\frac{\overline{A}(m^2_K)}{F^2}
+\left(\frac{1}{6}+\frac{1}{2}g^2\right)\frac{\overline{A}(m^2_\eta)}{F^2}\,.
\ea
$F_{B_s\to\pi}$ vanishes due to the possible flavour quantum numbers.

In all the transitions we obtain, as predicted by our arguments, the same
coefficients for the relativistic theory. I.e.
\be
\label{Bfpo}
f_{+/0}(q^2) = f_{+/0}^{\mathrm{Tree}}(q^2) F_{B\to M}\,.
\ee
The correction
is also the same for the formfactors $f_0$ and $f_+$ or for $f_v$ and $f_p$ in
all the cases. Notice that (\ref{Bp:finalresult}) is also in
agreement with the results in two-flavour HPChPT of \cite{Bijnens:2010ws}

The chiral logarithms for both form factors are always the same in these
decays as can be seen in (\ref{Bfpo}) and (\ref{Bfpv}). This was also
already the case for the $K_{\ell3}$ formfactors in HPChPT
\cite{Flynn:2008tg} and we noticed it as well in \cite{Bijnens:2010ws}.
It cannot simply be something like heavy quark symmetry since it is not valid
at the endpoint, see below and \cite{Falk:1993fr,Becirevic:2003ad}.
This would also not be a valid reason for the $K_{\ell3}$ case.
An alternative explanation would be if something similar to Low's theorem for
electromagnetic soft corrections holds. Low's theorem
states that the amplitude for the
process with Bremsstrahlung is proportional to the amplitude without
Bremsstrahlung by a factor depending only on the external legs.
A corresponding  result holds for the infrared logarithms in virtual
photon diagrams.
But, if here there was only dependence on the external legs,
we obtain the relation
\be
F_{B\to K}-F_{B\to\eta}-F_{B_s\to K}+F_{B_s\to\eta}=0\,.
\ee
Inspection of the results in (\ref{Bse:finalresult}) show that this
is not satisfied.
The same argument would have predicted that the chiral logarithms in
$F_V^\pi(s)$ and $F_S^\pi(s)$ of (\ref{FS:oneloop}) are the same which is
again clearly not the case.

\subsection{Comparison with experiment}

\begin{figure}
\begin{minipage}{0.49\textwidth}
\includegraphics[angle=270, width=0.99\textwidth]{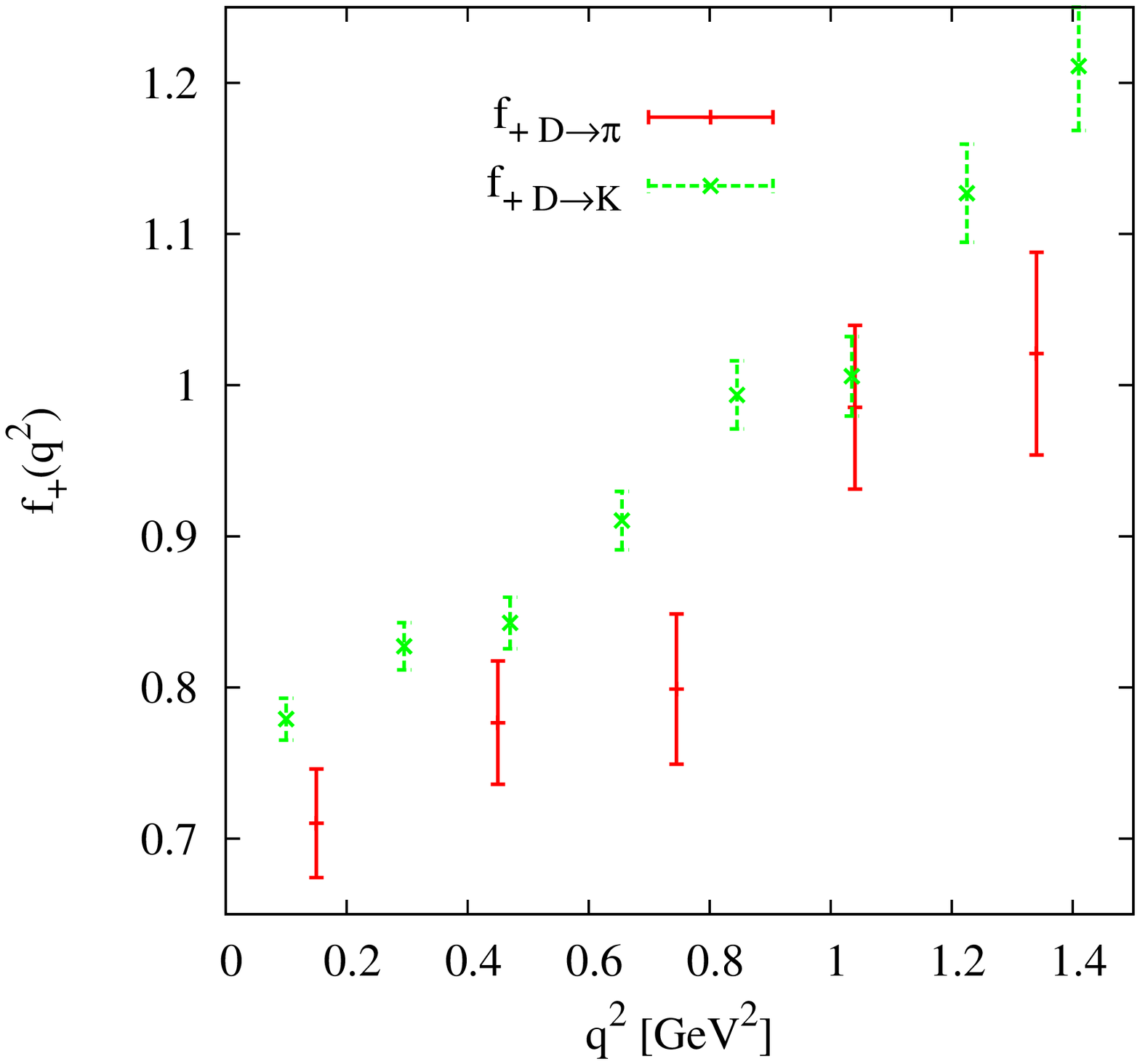}
\end{minipage}
\begin{minipage}{0.49\textwidth}
\includegraphics[angle=270, width=0.99\textwidth]{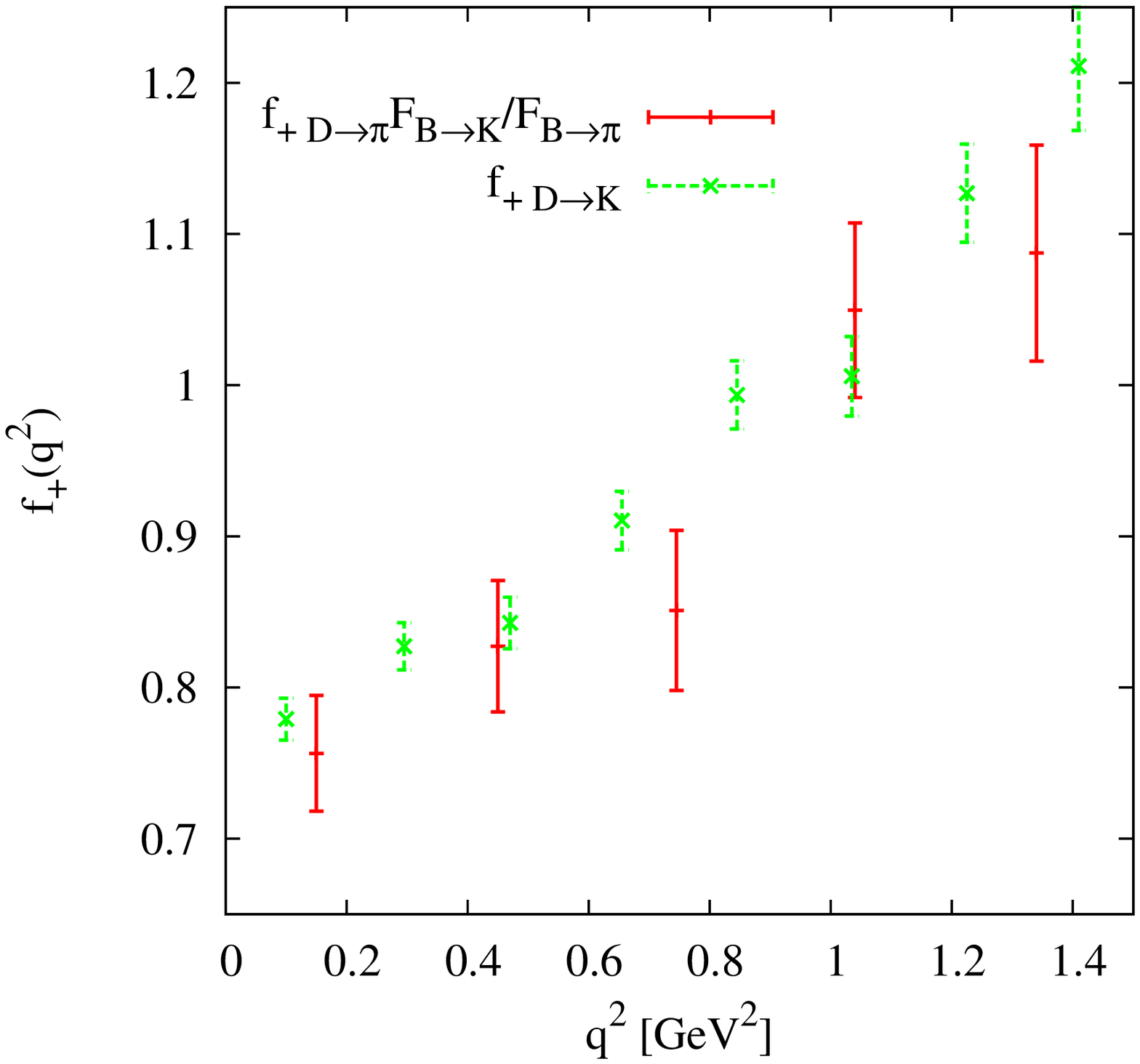}
\end{minipage}
\caption{\label{cleoplot} The measurements of CLEO \cite{:2008yi} for the
  formfactor $f_+(q^2)$ for the $D\rightarrow\pi$ and the $D\rightarrow K$
  semileptonic decays. In the two plots we have divided by the values
  of the CKM
  matrix elements $|V_{cd}|=0.2253$ and $|V_{cs}|=0.9743$
  respectively. On the left we plot only the formfactors without
  corrections, while on the right we plot each of the two members of
  (\ref{relplot}).}  
\end{figure}

We did not find any lattice data published in a form that allows us to test
the chiral logarithms in (\ref{Bse:finalresult}). However, there are
published data on the formfactors in $D\to\pi$ and $D\to K$ semileptonic
decays. The most precise data come from CLEO.
In \cite{:2008yi} are reported
the data points of $f_+(q^2)|V_{cd}|$ for $D^{+(0)}\rightarrow\pi^{0(+)}$
decays and
of $f_+(q^2)|V_{cs}|$ for
$D^{0(+)}\rightarrow K^{+(0)}$ decays.
We can then use the known value for the Cabibbo angle to get at 
the form factors. We used the PDG value for $\sin\theta_C=0.2253$
\cite{Nakamura:2010zzi}
 to obtain $|V_{cd}|=\sin\theta_C=0.2253$
 and $|V_{cs}|=\cos\theta_C=0.9743$.
In Fig.~\ref{cleoplot} on the left-hand-side
we plot the CLEO data for both $D\rightarrow \pi$
and $D\rightarrow K$
decays. We included only the $D^0\rightarrow\pi^+(K^+)$ data. A similar
study can be done using the $D^+\rightarrow\pi^0(K^0)$ since they give basically
the same data points as isospin symmetry dictates.

The following
relation should approximately hold using (\ref{Bfpo}) and the fact that the
lowest order result is the same.
\ba\label{relplot}
{f_{+\,D\rightarrow K}(q^2)} = {f_{+\,D\rightarrow  \pi}(q^2)}
     \frac{F_{D\rightarrow K}}{F_{D\rightarrow\pi}}\,,
\ea
where $F_{D\rightarrow\pi(K)}$ are the logarithmic corrections due to
loop diagrams quoted in  (\ref{Bse:finalresult}).
The corrections to relation (\ref{relplot}) are mainly due to
higher order terms i.e. $\mathcal{O}(m^2_M)$ without logarithms. We expect
these corrections to be about $10\%$.
The value of $g^2$, which enters through the
$F_{D\rightarrow\pi(K)}$ of (\ref{relplot}) is set
to 0.44 \cite{deDivitiis:1998kj}. 
However it does not affect the plots since the coefficients of the chiral
logarithms proportional to $g^{2}$ are the same for the two decays. 
The scale of renormalization $\mu$ is set to
$\Lambda_{\rm ChPT}\approx 1\,\,{\rm GeV}$.
In the right plot of
Fig.~\ref{cleoplot} the $f^+_{D\rightarrow K}$ formfactor
almost overlaps the $f^+_{D\rightarrow \pi}$ one once the logarithmic
corrections are taken into account as (\ref{relplot}) indicates. 
By comparing the left with the right plots in
Fig.~\ref{cleoplot} it is clear that our chiral logarithms compensate
for the differences.
Notice also that the $F_{D\rightarrow\pi(K)}$ contributes to a
good $30\%$ of the total formfactor but the total
correction in the ratio is much smaller.
The terms which depend on $g^2$ also cancel out in the ratio.
(\ref{relplot}) holds in principle both for $q^2\ll q^2_{\rm max}$
and at the endpoint $q^2\approx
q^2_{\rm max}$. It should be kept in mind that the endpoint
has quite different logarithms which are given below.
The $q^2_{\rm max}$  are rather
different in the two decays, being $q^2_{\rm max}
\approx(1.86-0.49)^2=1.88\,\,{\rm GeV}^2$  for the $K$
channel while $q^2_{\rm max} \approx(1.86-0.14)^2=2.9\,\,{\rm GeV}^2$
for the $\pi$ channel. Therefore making a similar comparison at large $q^2$ is
in practice not possible. This is the reason why in Fig.~\ref{cleoplot} we
stopped at $q^2\approx 1.5~{\rm GeV}^2$, the rightmost point is
already rather close to the endpoint for $D\to K$.

\subsection{Chiral logarithms at the endpoint}
\label{Bl3:endpoint}

At the endpoint HPChPT is not valid but standard HMChPT is.
The $B\to K$ formfactors were calculated in \cite{Falk:1993fr}
and the $B\to\pi,K$ in \cite{Becirevic:2003ad}.
The latter paper also discussed them in the partially quenched case. 
We do not show diagram by diagram results, these can be partly found
in \cite{Falk:1993fr,Becirevic:2003ad}.
Here we only quote the final results
but we also calculate the results for the $B\to\eta$ transitions.

Again the results in this limit must give the same outcome for the two
theories, since one is the relativistic limit of the
other. So this is another check of the validity of our relativistic
theory. Notice that we are performing a three-flavour calculation and
thus there are three light masses entering into the loop-functions,
i.e. $m_\pi$, $m_K$ and $m_\eta$. This complicates the structures of
the functions involved and therefore of the non-analyticities arising. 
For this reason a few
more loop functions are also needed in the relativistic formalism compared to
the two-flavour case \cite{Bijnens:2010ws}. They have been
reported in the appendix.
We present the results at  $q^2 = q^2_{\rm max}$ for each transition using
\be
f_p (q^2_{\rm max}) = f_p^{\rm Tree} (q^2_{\rm max})F^p_{B\to M}\,,
\qquad
f_v (q^2_{\rm max}) = f_v^{\rm Tree} (q^2_{\rm max})F^v_{B\to M}\,.
\ee
The relativistic theory correctly reproduces all these results
provided that the substitutions
$f^{\rm Tree}_v(v\cdot p)\rightarrow f^{\rm Tree}_0(q^2)$
and $f^{\rm Tree}_p(v\cdot p)\rightarrow f^{\rm Tree}_+(q^2)$ are performed.
\ba
\label{Bp:qsqmax}
F^p_{B\to\pi}&=&
1+\left(\frac{3}{8}+\frac{43}{24}g^2\right)\frac{\overline{A}(m^2_\pi)}{F^2}
+\left(\frac{1}{4}+\frac{9}{4}g^2-\frac{m^2_\pi}{m^2_K}g^2\right)\frac{\overline{A}(m^2_K)}{F^2}
\nonumber\\&&
+\left(\frac{1}{24}+\frac{11}{24}g^2-\frac{2}{9}\frac{m^2_\pi}{m^2_\eta}g^2\right)\frac{\overline{A}(m^2_\eta)}{F^2}
+2g^2\frac{\left(m^2_\pi-m^2_K\right)}{ F^2}\mathcal{F}\left(\frac{m_\pi}{m_K}\right)
\nonumber\\&&
+\frac{4}{9}g^2\frac{\left(m^2_\pi-m^2_\eta\right)}{ F^2}\mathcal{F}\left(\frac{m_\pi}{m_\eta}\right)
\nonumber\\
F^v_{B\to\pi}&=&
1+
\left(\frac{11}{8}+\frac{9}{8}g^2\right)\frac{\overline{A}(m^2_\pi)}{F^2}
+\left(-\frac{1}{4}+\frac{3}{4}g^2+\frac{m^2_\pi}{m^2_K}\right)\frac{\overline{A}(m^2_K)}{F^2}
\nonumber\\&&
+\left(\frac{1}{24}+\frac{1}{8}g^2\right)\frac{\overline{A}(m^2_\eta)}{F^2}
-2\frac{m^2_\pi}{F^2}\mathcal{F}\left(\frac{m_\pi}{m_K}\right),
\\
\label{BK:qsqmax}
F^p_{B\to K}&=&
1+\frac{9}{8}g^2\frac{\overline{A}(m^2_\pi)}{F^2}
+\left(\frac{1}{2}+\frac{7}{4}g^2\right)\frac{\overline{A}(m^2_K)}{F^2}
+\left(\frac{1}{6}+\frac{23}{24}g^2-\frac{5}{9}\frac{m^2_K}{m^2_\eta}g^2\right)\frac{\overline{A}(m^2_\eta)}{F^2}
\nonumber\\&&
+\frac{10}{9}g^2\frac{\left(m^2_K-m^2_\eta\right)}{F^2}\mathcal{F}\left(\frac{m_K}{m_\eta}\right),
\nonumber\\
F^v_{B\to K}&=&
1+\frac{9}{8}g^2\frac{\overline{A}(m^2_\pi)}{F^2}
+\left(\frac{3}{2}+\frac{3}{4}g^2\right)\frac{\overline{A}(m^2_K)}{F^2}
+\left(-\frac{1}{3}+\frac{1}{8}g^2+\frac{m^2_K}{m^2_\eta}\right)\frac{\overline{A}(m^2_\eta)}{F^2}
\nonumber\\&&
-2\frac{m^2_K}{F^2}\mathcal{F}\left(\frac{m_K}{m_\eta}\right),
\\
\label{Beta:qsqmax}
F^p_{B\to\eta}&=&
1+
\left(\frac{3}{8}+\frac{33}{8}g^2-2\frac{m_\pi^2}{m_\eta^2}g^2\right)\frac{\overline{A}(m^2_\pi)}{F^2}
+\left(\frac{1}{4}+\frac{5}{4}g^2-\frac{1}{3}\frac{m^2_\eta}{m^2_K}g^2\right)\frac{\overline{A}(m^2_K)}{F^2}
\nonumber\\&&
+\left(\frac{1}{24}+\frac{17}{72}g^2\right)\frac{\overline{A}(m^2_\eta)}{F^2}+4g^2\frac{\left(m^2_\eta-m^2_\pi\right)}{F^2}\mathcal{F}\left(\frac{m_\eta}{m_\pi}\right)
\nonumber\\&&
+\frac{2}{3}g^2\frac{\left(m^2_\eta-m^2_K\right)}{F^2}\mathcal{F}\left(\frac{m_\eta}{m_K}\right),
\nonumber\\
F^v_{B\to\eta}&=&
1+\left(\frac{3}{8}+\frac{9}{8}g^2\right)\frac{\overline{A}(m^2_\pi)}{F^2}
+\left(-\frac{5}{4}+\frac{3}{4}g^2+3\frac{m^2_\eta}{m^2_K}\right)\frac{\overline{A}(m^2_K)}{F^2}
\nonumber\\&&
+\left(\frac{1}{24}+\frac{1}{8}g^2\right)\frac{\overline{A}(m^2_\eta)}{F^2}
-6\frac{m^2_\eta}{F^2}\mathcal{F}\left(\frac{m_\eta}{m_K}\right),
\\
F^p_{B_s\to K}&=&
1+\left(\frac{3}{8}+\frac{9}{4}g^2-\frac{3}{2}\frac{m^2_K}{m^2_\pi}g^2\right)\frac{\overline{A}(m^2_\pi)}{F^2}
\nonumber\\&&
+\left(\frac{1}{4}+2g^2\right)\frac{\overline{A}(m^2_K)}{F^2}
+\left(\frac{1}{24}+\frac{7}{12}g^2-\frac{1}{18}\frac{m^2_K}{m^2_\eta}g^2\right)\frac{\overline{A}(m^2_\eta)}{F^2}
\nonumber\\&&
+3g^2\frac{(m^2_K-m^2_\pi)}{F^2}\mathcal{F}\left(\frac{m_K}{m_\pi}\right)
+\frac{1}{9}\frac{(m^2_K-m^2_\eta)}{F^2}\mathcal{F}\left(\frac{m_K}{m_\eta}\right),
\nonumber\\
F^v_{B_s\to K}&=&
1+\left(-\frac{3}{8}+\frac{3}{2}\frac{m^2_K}{m^2_\pi}\right)\frac{\overline{A}(m^2_\pi)}{F^2}
+\left(\frac{3}{4}+\frac{3}{2}g^2\right)\frac{\overline{A}(m^2_K)}{F^2}
\nonumber\\&&
+\left(-\frac{5}{24}+\frac{1}{2}g^2+\frac{1}{2}\frac{m^2_K}{m^2_\eta}\right)\frac{\overline{A}(m^2_\eta)}{F^2}
-3\frac{m^2_K}{F^2}\mathcal{F}\left(\frac{m_K}{m_\pi}\right)
-\frac{m^2_K}{F^2}\mathcal{F}\left(\frac{m_K}{m_\eta}\right),
\\
\label{Bseta:qsqmax}
F^p_{B_s\to\eta}&=&
1
+\left(\frac{1}{2}+4g^2-\frac{5}{3}\frac{m^2_\eta}{m^2_K}g^2\right)\frac{\overline{A}(m^2_K)}{F^2}
+\left(\frac{1}{6}+\frac{17}{18}g^2\right)\frac{\overline{A}(m^2_\eta)}{F^2}
\nonumber\\&&
+\frac{10}{3}\frac{(m^2_\eta-m^2_K)}{F^2}\mathcal{F}\left(\frac{m_\eta}{m_K}\right),
\nonumber\\
F^v_{B_s\to\eta}&=&
1
+\left(-1+\frac{3}{2}g^2+3\frac{m^2_K}{m^2_\pi}\right)\frac{\overline{A}(m^2_K)}{F^2}
+\left(\frac{1}{6}+\frac{1}{2}g^2\right)\frac{\overline{A}(m^2_\eta)}{F^2}
\nonumber\\&&
-6\frac{m^2_\eta}{F^2}\mathcal{F}\left(\frac{m_\eta}{m_K} \right),
\ea
with 
\ba\label{F}
\mathcal{F}\left(\frac{m_1}{m_2}\right)=\left\{\begin{array}{ll}
-\frac{1}{(4\pi)^2}\frac{\sqrt{m^2_2-m^2_1}}{m_1}\left[\frac{\pi}{2}-\arctan{\left(\frac{m_1}{\sqrt{m^2_2-m^2_1}}\right)}\right]&
m_1\leq m_2\\
\frac{1}{(4\pi)^2}\frac{\sqrt{m^2_1-m^2_2}}{m_1}\tanh^{-1}{\left(\frac{\sqrt{m^2_1-m^2_2}}{m_1}\right)}&m_1\geq m_2
\end{array} \right.
\ea
Our results agree with the earlier published ones in 
\cite{Falk:1993fr,Becirevic:2003ad}.

\section{$B\rightarrow D$ transition}
\label{BtoD}

\subsection{Definition of formfactors}
\label{BD:formalism}

In this section we present the formalism involved in the calculation of the
$B\rightarrow D$ formfactor.
The matrix element for this decay is
\ba\label{BD:QCDformfact1}
\left< D(p') \left|\bar{b} \gamma_\mu c\right|B(p)\right>
&=&(p+p')_\mu \tilde{f}_+(q^2)+ (p-p')_\mu \tilde{f}_-(q^2)
\ea
where $q^\mu$ is the momentum transfer $q^\mu=p-p'$.

To perform the calculation in HMChPT we need the hadronic current
corresponding to the one of QCD:
\ba
\label{BD:matching}
\bar{b} \gamma_\mu c \rightarrow \Tr{X(v,v') \bar{H}(v') \gamma_\mu H(v)}
\ea
where $v$, $v'$ are the fixed four-velocities of the $B$ and $D$ hadron
respectively, while $X(v,v')$ is the most general bispinor constructed
starting from the invariants $v$ and $v'$. As explained in
\cite{Heavyquarkbook}, spin symmetry for heavy quarks constrains $X$ to be a
scalar function $-\xi(v\cdot v')$, called the Isgur-Wise
function \cite{Isgur:1989ed}.
The variable $v\cdot v'$ is of special importance. It can
be related to $q^2$ through the relation
\be
w\equiv v\cdot v'=\frac{m^2_B+m^2_D-q^2}{2 m_B m_D}\,.
\ee
The allowed kinematic range is thus $0\leq w-1 \leq
\frac{(m_B-m_D)^2}{2m_Bm_D}$. $w$ is a measure of what is
the momentum transfer to the light degrees of freedom i.e. it gives us an
indication of the range of applicability of HMChPT. The light degrees of
freedom have momentum of order $\Lambda_{\rm QCD}v^{(')}$, thus the momentum
transfer to the light system is 
$q^2_{\rm light}\approx\left(\Lambda_{\rm QCD}v-\Lambda_{\rm QCD}v'\right)^{2}=2\Lambda^2_{\rm QCD}(1-w)$. 
HMChPT can be applied as far as $q^2_{\rm light}\ll m^2_{b,c}$
which means on the scale $w\approx 1$ (region of zero recoil
or near the endpoint) \cite{Randall:1993qg}.
The matrix element in HMChPT is
\ba\label{BD:HMChPTformfact}
\left< D(v') \left|\bar{b} \gamma_\mu c\right|B(v)\right>_{\rm HMChPT}
&=&(v+v')_\mu h_{+}(w).
\ea
Evaluating explicitely the trace in (\ref{BD:matching})  it is easy to
obtain $h_{+}(w)=\xi(w)$ at leading order. It can be also shown that heavy
flavour symmetry implies $\xi(1)=1$ \cite{Isgur:1989ed,Heavyquarkbook}.
The result that one single formfactor is enough
to describe the matrix element of (\ref{BD:HMChPTformfact}) can also be
achieved using the helicity formalism for counting the number of independent
amplitudes \cite{Heavyquarkbook,Politzer:1990ps}.
To compare with the results of HMChPT it is convenient to reparametrize the
matrix element of QCD defined in (\ref{BD:QCDformfact1}) as
\ba\label{BD:QCDformfact2}
\frac{\left< D(p') \left|\bar{b} \gamma_\mu c\right|B(p)\right>}{\sqrt{m_B m_D}}
&=&(v+v')_\mu h_+(w)+ (v-v')_\mu h_-(w)\,.
\ea
where the formfactors $h_\pm(w)$  are linear combinations of 
$\tilde{f}_\pm(q^2)$.
Comparing (\ref{BD:HMChPTformfact}) and (\ref{BD:QCDformfact2}) it is
straightforward to see that, at leading order in $1/m_{heavy}$,
$h_{+}(w)$ must be the same formfactor in the two formalisms
and that $h_-(w)=0$.

To perform the calculation in the relativistic framework we need the $J^L_\mu$
current responsible for the $B\rightarrow D$ transition, analogous
to the one in (\ref{relleftcurr}). Therefore we write down all the possible
independent and chiral-invariant
operators that respect also heavy quark symmetries. 
They must contain interactions of the kind $BD$ or $B^*D^*$. The first
one is needed for the tree-level diagram (1) in Fig.~\ref{fig:BDdiagram},
while the second for the
one-loop (2) in Fig.~\ref{fig:BDdiagram}. Thus the current is
\ba
\label{BD:relleftcurr}
J^L_\mu&=&X_1\left(-tD^\dagger\nabla_\mu B+t\nabla_\mu D^\dagger
B\right)+X_2\left(tD^{*\,\dagger }_{\alpha} \nabla_\mu
B^{*\,\alpha}-t\nabla_\mu D^{*\,\dagger }_{\alpha} B^{*\,\mu}\right)
\nonumber\\&&
+X_3\left(-t\nabla^\alpha D^{*\,\dagger }_{\alpha} B^*_{\mu}
+t D^{*\,\dagger}_{\mu}\nabla^\alpha B^*_{\alpha}
+t\nabla^\alpha D^{*\,\dagger }_{\mu} B^*_{\alpha}
-t D^{*\,\dagger }_{\alpha}\nabla^\alpha B^*_{\mu}\right)\,
\ea
where $X_1,X_2,X_3$ are effective couplings and the
spurion $t$ is now a singlet under the chiral $SU(n)_L\times SU(n)_R$ 
symmetry since $\overline b\gamma_\mu c$ is a singlet.
Heavy quark symmetry implies furthermore that $X_1=X_2=X_3$.
From (\ref{BD:relleftcurr}) it is easy to construct the vector current $J^V_\mu$
causing the decay.

Before concluding this section we stress once
more that the zero recoil region is the only one where HMChPT is in principle
applicable, as shown by \cite{Randall:1993qg}. 
This does not mean that it is not possible
to extend the effective theory outside that range to calculate the infrared
singularities. Indeed exactly the same arguments
applied to $B\rightarrow \pi$ semileptonic decays go through for the
$B\rightarrow D$ case as well, thus HPChPT can be used. 
As a matter of fact there have been already confirmations of how well the
effective theory can do when $w-1\gg 0$ (see for example Fig~2.5 in
\cite{Heavyquarkbook}). The use
of HPChPT justify those results.

\subsection{Chiral logarithms}
\label{BD:ChLogs}

We now present the results for the $B_{(s)}\rightarrow D_{(s)}$
semileptonic decay.
The results in two-flavour HMChPT at zero recoil ($w=1$) can be found in
\cite{Randall:1993qg}. The three-flavour extension  has been calculated
in \cite{Jenkins:1992qv} and \cite{Boyd:1995pq}.
The result at one loop and leading order in $1/m_{B}$ and $1/m_{D}$ is
\be
\label{xi:oneloop}
h_{+}(w)=\xi(w)\left[1+\frac{g^2}{F^2}\left(\frac{3}{2}\overline{A}(m^2_\pi)
+\overline{A}(m^2_K)
+\frac{1}{6}\overline{A}(m^2_\eta)\right)\left(r(w)-1\right)\right],
\ee
and for the $B_s\to D_s$ transition it is
\be
\label{xis:oneloop}
h_{+}(w)=\xi(w)\left[1+\frac{g^2}{F^2}\left(
2\overline{A}(m^2_K)
+\frac{2}{3}\overline{A}(m^2_\eta)\right)\left(r(w)-1\right)\right],
\ee
where
\be
r(w)=\frac{1}{\sqrt{w^2-1}}\log{(w+\sqrt{w^2-1})},
\ee
and $r(1)=1$ so that the chiral logarithms cancel at zero recoil.
While in \cite{Randall:1993qg} it has been clearly stated that the
calculation is valid only at the zero recoil point, the authors
of \cite{Jenkins:1992qv} and
\cite{Boyd:1995pq} present the
result for the Isgur-Wise function in the whole energy range, but no explicit
arguments why it should be valid are given there.
The arguments of HPChPT given before imply that the formula given there
are indeed valid in the whole energy regime $0\leq w-1 \leq
\frac{(m_B-m_D)^2}{2m_Bm_D}\approx1.6$.
Note that here the correction is not a simple chiral logarithm as in
the previous cases but there is a strong dependence on $w$ and the
result connects smoothly to the endpoint region.

We checked that our relativistic formulation gives the same result
as \cite{Boyd:1995pq}.
The result up to one-loop reads
\ba
\label{h+:relth}
h_+(w)\hspace*{-0.2cm}&=&\hspace*{-0.2cm}
\frac{X_1}{\sqrt{2}}\left[1+\frac{g^2}{F^2}\left(
\frac{3}{2}\overline{A}(m^2_\pi)+
\overline{A}(m^2_K) +
\frac{1}{6}\overline{A}(m^2_\eta)\right)\left(1-2m_Bm_D\tilde{C}(m_D^2,m_B^2,q^2)\right)\right],\nonumber\\
h_-(w)\hspace*{-0.2cm}&=&\hspace*{-0.2cm}0,
\ea
where $\tilde{C}(m^2_D,m^2_B,q^2)$ comes from the
three-point function $C(m^2,m^2_B,m^2_D,m^2_B,q^2,m^2_D)$ which is needed
to evaluate the loop diagram in Fig.~\ref{fig:BDdiagram}. 
In (\ref{BDC}) in the appendix we define
the function $\tilde{C}(m^2_D,m^2_B,q^2)$ and show that
$\tilde{C}(m^2_D,m^2_B,q^2)=r(w)/(2m_Bm_D)$. We also agree with the 
$B_s\to D_s$ result of \cite{Boyd:1995pq}.

Comparing (\ref{h+:relth}) with (\ref{xi:oneloop}) it is straightforward to see
that the two formalisms give the same results as foreseen.
Notice that now we do not need to distinguish the two limits
$w-1\approx0$ and $w- 1\gg0$: the function $r(w)$ describes
the whole energy range.

Note that the reults here assume that there are no other nearby states,
see e.g. the discussion in \cite{Eeg:2007ha}.
\begin{figure}
\begin{center}
\includegraphics{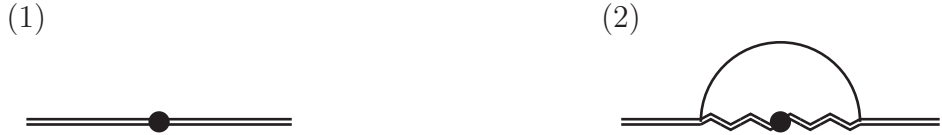}
\end{center}
\caption{The diagrams contributing to the $B\rightarrow D$ transition up to
  one-loop. Notation is the same as in Fig.~\ref{fig:treelevel}. The double
  lines at
  the left of the insertion of the current are always $B$ mesons, while the
  ones in the right are $D$ mesons.}
\label{fig:BDdiagram}
\end{figure}

\section{Conclusions}

In the paper we have extended HPChPT to several processes. 
First we calculated the three-flavour results for the charged pion and kaon
electromagnetic formfactor and the two-flavour result for the pion
vector and scalar formfactor. The latter have then been used to check
the underlying arguments of HPChPT in a two-loop setting.

Using the three-flavour
spontaneous symmetry breaking pattern we could explicitely evaluate
the dependence on the light meson masses for the $B\rightarrow\pi$
formfactors in addition to our earlier
two-flavour results \cite{Bijnens:2010ws}.
We could also extend the theory to other transitions as the
$B\rightarrow K$ and $B\rightarrow\eta$ transitions and the corresponding
$B_s$ transitions. The corrections are of the expected size of about 30\%.
An unexplained feature of our results is that the two 
possible formfactors have always the same chiral logarithm and we ruled out two
possible explanations.
A comparison with the experimental data for the $D\to\pi,K$ transition
formfactors has also performed. It shows that the corrections obtained
go in the right direction and are sizable. We have reproduced the known results
at the endpoints and added these for the transitions to $\eta$.

Finally, we justified
and reproduced already known results for the formfactors of the $B\rightarrow
D$ transition at one loop.

Further investigations in this framework are
desiderable, since they could significantly improve the chiral extrapolations
of the lattice data. In particular it could be very useful to develop the same
approach also for Partially Quenched ChPT. As stated above we expect that a
formalism with an explicit power counting can be formulated along the
lines of SCET.

\section*{Acknowledgments}

This work is supported in part by the European Community-Research
Infrastructure Integrating Activity ``Study of Strongly Interacting Matter'' 
(HadronPhysics2, Grant Agreement n. 227431)
and the Swedish Research Council grants 621-2008-4074 and 621-2008-4252.
This work heavily used FORM
\cite{Vermaseren:2000nd}.

\appendix
\setcounter{equation}{0}
\renewcommand{\theequation}{\Alph{section}.\arabic{equation}}

\section{Expansion of the needed loop integrals}
\label{appendix}

We collect here the one-loop
functions and their expansions, used to evaluate the diagrams in
Fig.~\ref{fig:oneloop} and in Fig.~\ref{fig:BDdiagram} in the framework
of the relativistic
theory of Sect.~\ref{relth}. 
Much of what is written here is also present in the appendix of
\cite{Bijnens:2010ws}. We need the
one-, two- and three-point
functions defined as ($d=4-2\epsilon$)
\ba
\label{A}
A(m^2_1)&=&\frac{1}{i}\int{\frac{d^dk}{(2\pi)^d}\frac{1}{k^2-m^2_1}},\\
\label{B}
B(m^2_1,m^2_2,p^2)&=&
\frac{1}{i}\int{\frac{d^dk}{(2\pi)^d}\frac{1}{(k^2-m^2_1)((p-k)^2-m^2_2)}},\\
\label{C}
C(m^2_1,m^2_2,m^2_3,p^2_1,p^2_2,q^2)&=&\frac{1}{i}
\int{\frac{d^dk}{(2\pi)^d}
\frac{1}{(k^2-m^2_1)((k-p_1)^2-m^2_2)((k-p_1-p_2)^2-m^2_3)}},\nonumber\\
\ea
with $q^2 =(p_1+p_2)^2$.
Two- and three-point functions with extra powers of momenta in the
numerator contribute too. They are defined similarly
and the explicit definitions can be found in
\cite{Bijnens:2002hp}. All these
functions can be rewritten in terms of (\ref{A}), (\ref{B}) and (\ref{C})
\cite{Passarino:1978jh}. 
The finite parts of $A(m^2_1)$ and $B(m^2_1,m^2_2,q^2)$
are using the standard ChPT subtractions \cite{GL1,GL2,'tHooft:1978xw}
\ba
\label{Abar}
\overline{A}(m^2_1)&=&
 -\frac{m^2_1}{16\pi^2}\log{\left(\frac{m^2_1}{\mu^2}\right)},\\
\label{Bbar}
\overline{B}(m^2_1,m^2_2,q^2)&=&
\frac{1}{16\pi^2}\left[-1-\int^1_0dx
\log{\left(\frac{m_1x+m_2(1-x)-x(1-x)q^2}{\mu^2}\right)}\right].
\ea

As far as regards the $B$ transitions
to a light pseudoscalar meson, the three-point function
$C(m^2_1,m^2_2,m^2_3,p^2_1,p^2_2,q^2)$
always depends on the masses as $(m_1^2,M^2,M^2,M^2,m_2^2,q^2)$ where $m_1$ is
the mass of the light meson in the loop, $m_2$ is the mass of the light
external meson and $M=m_B$. It can be
rewritten using Feynman parameters $x$, $y$
\ba
\label{Cbar}
\lefteqn{
C(m^2_1,M^2,M^2,M^2,m^2_2,q^2)=
-\frac{1}{16\pi^2}\int^1_0dx\int^{1-x}_0 dy\,\times}
\nonumber\\&&
\left[m^2_1(1-x-y)+m^2_2(-y+y^2)
+M^2(x+y)^2+(q^2-M^2-m^2_2)(-y+y(x+y))\right]^{-1}.
\nonumber\\&&
\ea
In order to find the appropriate chiral logarithms we expanded (\ref{Bbar}) and
(\ref{Cbar}) for small ratios $m^2/M^2$ . We quote only the terms of the
expansions containing non-analyticities in the light masses $m_i$.
First those only valid for  $q^2\ll q^2_\mathrm{max}$, i.e. away from
the endpoint:
\ba  
\overline{B}(m^2,M^2,q^2)& =&
 -\frac{1}{M^2-q^2}\overline{A}(m^2),
\\
\label{Cexp2}
C(m^2_1,M^2,M^2,M^2,m^2_2,q^2)&=&
\frac{1}{(M^2-q^2)}\frac{1}{16\pi}\frac{m_1}{M}
-\frac{1}{(M^2-q^2)^2}\overline{A}(m^2_1)\,.
\ea
The next ones are those relevant at the endpoint or for wavefunction
renormalization
\ba
\label{Bexp}
\bar{B}(m^2,M^2,M^2)& = &
-\frac{1}{16\pi}\frac{m}{M}+\frac{1}{16\pi}\frac{m^2}{M^2}
+\frac{1}{2M^2}\overline{A}(m^2),
\\
\bar{B}(m^2,M^2,m^2) &=& 0,\\
\label{Bbarmax}
\bar{B}(m^2_1,M^2,(M-m_2)^2)& =&
\frac{1}{M}\left[ 2m_2\mathcal{F}\left(\frac{m_1}{m_2}\right)
  -\frac{m_2}{m_1^2}\overline{A}(m^2_1)\right]
\nonumber\\
&&+\frac{1}{M^2}\left[ 
  3m_2^2\mathcal{F}\left(\frac{m_1}{m_2}\right)+
  \overline{A}(m^2_1)\left(\frac{1}{2}
  -\frac{3}{2}\frac{m_2^2}{m_1^2}\right)\right].
\ea
The function $\mathcal{F}(m_1/m_2)$ was defined in (\ref{F}). The expansion
in (\ref{Bbarmax}) holds in both the cases $m_2\lessgtr m_1$ and also for $m_1=m_2$ where it correctly reduces to the expansion reported in the appendix
of \cite{Bijnens:2010ws}.

The expansions of the three-point functions at $q^2_\mathrm{max}$ are a bit more
involved. The reason is that the reduction formulas present a singularity at
$q^2_\mathrm{max}=(M-m_2)^2$ for $m^2_2=0$.
Furthermore we need to distinguish different cases depending on the
$m_1$ and $m_2$ appearing in the arguments of the three-loop functions.
We use again the same technique used in \cite{Bijnens:2010ws}. We expand each
of the functions directly from the Feynman parameter integral, without
first rewriting them in terms of (\ref{A}), (\ref{B}) and (\ref{C}).
To do this one rewrites the integral in (\ref{Cbar}) using $z=x+y$
as
\ba
\lefteqn{C(m^2_1,M^2,M^2,M^2,m^2_2,(M-m_2)^2)=
-\frac{1}{16\pi^2}\int^1_0dz\int^{z}_0dy \times}&&
\nonumber\\&&
\frac{1}
{\left[M^2z^2+m^2_1+2m_2My+\left(-m^2_1z+m^2_2(-y+y^2)-2m_2Myz\right)\right]}.
\ea
The part in the denominator in brackets is always suppressed by at least $m/M$
compared to the first three terms for all values of $z$ and $y$ and we can
thus expand in it. The remaining integrals can be done with elementary means.
The expansions obtained are many and long, therefore we quote only those needed
and restricted to those terms where a infrared singularity
appears. We
do not quote terms like $1/(4\pi)^2m/M^3$, also non-analytic
for small $m$, because they always cancel in the final results
(\ref{Bp:qsqmax})-(\ref{Bseta:qsqmax}). The expansions read
\ba
\label{Cmax} 
\lefteqn{
C(m^2_1,M^2,M^2,M^2,m^2_2,(M-m_2)^2)=
\frac{1}{M^2}\left[-\frac{1}{2}\frac{1}{m_1^2}\overline{A}(m^2_1)
  +\mathcal{F}\left(\frac{m_1}{m_2}\right)\right]
}&&
\nonumber\\&&
\left.
+\frac{1}{M^3}\left[\mathcal{F}\left(\frac{m_1}{m_2}\right)m_2\right.
-\frac{1}{2}\frac{m_2}{m_1^2} \overline{A}(m^2_1)\right]
+\frac{1}{M^4}\left[-\frac{1}{2}\frac{m_2^2}{m_1^2}\overline{A}(m^2_1)+\mathcal{F}\left(\frac{m_1}{m_2}\right)\left(m_2^2+\frac{3}{8}m_1^2 \right)\right],
\nonumber\\
\lefteqn{
   C_{11}(m^2_1,M^2,M^2,M^2,m^2_2,(M-m_2)^2)=\frac{1}{M^3}\left(-\mathcal{F}\left(\frac{m_1}{m_2}\right)m_2+\frac{1}{2}\frac{m_2}{m_1^2}\overline{A}(m^2_1)\right)
}&&
\nonumber\\&&
+ \frac{1}{M^4}\left[\left(- \frac{1}{2} + \frac{13}{12}\frac{m_2^2}{m_1^2}\right)\overline{A}(m^2_1)- \mathcal{F}\left(\frac{m_1}{m_2}\right)\left(\frac{13}{6}m_2^2 - \frac{1}{6}m_1^2 \right)\right],
\nonumber\\
\lefteqn{C_{12}(m^2_1,M^2,M^2,M^2,m^2_2,(M-m_2)^2)=
\frac{1}{M^3}\left(-\mathcal{F}\left(\frac{m_1}{m_2}\right)\left(
\frac{2}{3}m_2 + \frac{1}{3}\frac{m^2_1}{m_2} \right)+
\frac{1}{3}\frac{m_2}{m_1^2}\overline{A}(m_1)\right)
}&&
\nonumber\\&&
+\frac{1}{M^4}\left[\left( - \frac{1}{4} +
  \frac{5}{6}\frac{m^2_2}{m^2_1}\right)
  \overline{A}(m^2_1)-\mathcal{F}\left(\frac{m_1}{m_2}\right)\left(
  \frac{5}{3}m_2^2 + \frac{1}{3}m_1^2 \right)\right],
\nonumber\\
\lefteqn{
   C_{21}(m^2_1,M^2,M^2,M^2,m^2_2,(M-m_2)^2)=\frac{1}{M^4}\left[-\mathcal{F}\left(\frac{m_1}{m_2}\right)\left(-\frac{4}{3}m_2^2
     +\frac{1}{3}m_1^2\right)\right.
}&&
\nonumber\\&&
\left.+\overline{A}(m^2_1)\left(\frac{1}{2}-\frac{2}{3}\frac{m_2^2}{m_1^2}\right)\right],
\nonumber\\
\lefteqn{
   C_{22}(m^2_1,M^2,M^2,M^2,m^2_2,(M-m_2)^2)=
   \frac{1}{M^4}\left[\mathcal{F}\left(\frac{m_1}{m_2}\right)\left(\frac{4}{5}m_2^2
     + \frac{1}{15}m_1^2 + \frac{2}{15}\frac{m_1^4}{m_2^2}\right)\right.
}&&
\nonumber\\&&
\left.+\overline{A}(m^2_1)\left(\frac{1}{6}
  -\frac{2}{5}\frac{m_2^2}{m_1^2}\right)\right],
\nonumber\\
\lefteqn{
   C_{23}(m^2_1,M^2,M^2,M^2,m^2_2,(M-m_2)^2)=\frac{1}{M^4}\left[\mathcal{F}\left(\frac{m_1}{m_2}\right)m_2^2
     + \overline{A}(m^2_1)\left(\frac{1}{4}
     -\frac{1}{2}\frac{m_2^2}{m_1^2}\right)\right],
}
\nonumber\\
\label{C24max}
\lefteqn{
   C_{24}(m^2_1,M^2,M^2,M^2,m^2_2,(M-m_2)^2)=
\frac{1}{M^2}\left[-\mathcal{F}\left(\frac{m_1}{m_2}\right)\frac{1}{3}(m_2^2 - m_1^2 )\right.
}&&
\nonumber\\&&
\left.
+\overline{A}(m_1)\left(-\frac{1}{4} + \frac{1}{6}\frac{m^2_2}{m^2_1}\right)\right].
\ea
Setting the masses $m_1=m_2$  all the expansions
in (\ref{C24max}) coincide correctly with the ones reported in the
appendix of \cite{Bijnens:2010ws}. The function $\mathcal{F}(m_1/m_2)$ is the
one defined in (\ref{F}). Notice that it takes different forms depending if
$m_1\lessgtr m_2$. Furthermore for $m_1=m_2$ $\mathcal{F}(m_1/m_2)=0$.
The other three-point functions do not give any leading contribution.

We focus now on the semileptonic decay $B\rightarrow D$. The three-point
function entering in the loop diagram of Fig.~\ref{fig:BDdiagram} is
$C(m^2,M^2_1,M^2_2,M^2_1,q^2,M^2_2)$, where $m$ is the mass of the light meson
in the loop, $M_1=m_B$ and $M_2=m_D$. To expand it, similarly to what
has been done above, we first rewrite it in terms of the Feynman
parameters $x$, $y$
\ba
\label{BDCbar}
\lefteqn{
C(m^2,M^2_1,M^2_2,M^2_1,q^2,M^2_2)=-\frac{1}{16\pi^2}\int^1_0dx\int^{1-x}_0
dy\times}&&
\nonumber\\&&
\left[m^2(1-x-y)+x^2M^2_1+y^2M^2_2
+xy(M^2_1+M^2_2-q^2)\right]^{-1}.
\ea
The $m^2(x+y)$ term in (\ref{BDCbar}) is suppressed by at least one power of
$m$ so we can neglect it.
Setting $x=X/M_1$, $y=Y/M_2$ and $w=(M^2_1+M^2_2-q^2)/(2M_1M_2)$ the integral
becomes
\ba
C(m^2,M^2_1,M^2_2,M^2_1,q^2,M^2_2)=
-\frac{1}{16\pi^2}\frac{1}{M_1M_2}\int^{M_1}_0\hspace{-0.4cm}dX
\int^{M_2-\frac{M_2}{M_1}X}_0\hspace{-0.5cm}dY
\left[m^2+X^2+Y^2+2wXY\right]^{-1}.\nonumber
\ea  
Then we can perform another change of variable and set polar coordinates
$X=R\cos{\phi}$, $Y=R\sin{\phi}$:
\ba
C(m^2,M^2_1,M^2_2,M^2_1,q^2,M^2_2)=
-\frac{1}{16\pi^2}\frac{1}{M_1M_2}\int^{\pi/2}_0 \hspace{-0.5cm}d\phi
\int^{R_{\rm  max}}_0
\hspace{-0.4cm}dR R\left[m^2+R^2+2wR^2\sin{(2\phi)}\right]^{-1},\nonumber
\ea 
where the upper boundary is $R_{\rm
  max}=M_2/(\sin{\phi}+M_2/M_1\cos{\phi})$. We are interested in
isolating the infrared singularities. Those only arise from the lower
bound of the integral. 
Therefore, performing the integral in $dR$, we keep only the term
coming from the small $R$ region. However we checked explicitely that the large
$R$ region does not produce any soft singularity at the desired order. The
result for the integral in $R$ reads
\ba
C(m^2,M^2_1,M^2_2,M^2_1,q^2,M^2_2)=
\frac{1}{16\pi^2}\frac{1}{2M_1M_2}\int^{\pi/2}_0 \hspace{-0.3cm}d\phi
\log{\left(\frac{m^2}{\mu^2}\right)}\left[1+2w\sin{(2\phi)}\right]^{-1}+
\cdots,\nonumber
\ea 
where the ellipsis are the terms coming from the upper bound and $\mu$ is a
parameter with the dimension of a mass. The integral
in $d\phi$ can be done analitycally and after tedious calculations we arrive
to the final result
\ba\label{BDC}
C(m^2,M^2_1,M^2_2,M^2_1,q^2,M^2_2)&=&\frac{1}{16\pi^2}\frac{1}{2M_1M_2}\frac{1}{\sqrt{w^2-1}}\log{\left(w+\sqrt{w^2-1}\right)}\log{\left(\frac{m^2}{\mu}\right)}+\dots\nonumber\\
&=&\frac{1}{16\pi^2}\tilde{C}(M^2_2,M^2_1,q^2)\log{\left(\frac{m^2}{\mu^2}\right)}+\dots\,\,.
\ea


\begin{thebibliography}{99}

\bibitem{Weinberg0}
  S.~Weinberg,
  Physica {\bf A96 } (1979)  327.

\bibitem{GL1}
  J.~Gasser and H.~Leutwyler,
  Annals Phys.\  {\bf 158} (1984) 142.

\bibitem{GL2}
  J.~Gasser and H.~Leutwyler,
  Nucl.\ Phys.\  B {\bf 250} (1985) 465.

\bibitem{:2008yi}
  J.~Y.~Ge {\it et al.}  [CLEO Collaboration],
  Phys.\ Rev.\  D {\bf 79} (2009) 052010
  [arXiv:0810.3878 [hep-ex]].

\bibitem{Adam:2007pv}
  N.~E.~Adam {\it et al.}  [CLEO Collaboration],
  Phys.\ Rev.\ Lett.\  {\bf 99} (2007) 041802
  [arXiv:hep-ex/0703041].

\bibitem{Hokuue:2006nr}
  T.~Hokuue {\it et al.}  [Belle Collaboration],
  Phys.\ Lett.\  B {\bf 648} (2007) 139
  [arXiv:hep-ex/0604024].

\bibitem{Aubert:2006px}
  B.~Aubert {\it et al.}  [BABAR Collaboration],
  Phys.\ Rev.\ Lett.\  {\bf 98} (2007) 091801
  [arXiv:hep-ex/0612020].

\bibitem{Gamiz:2008iv}
  E.~Gamiz,
  PoS {\bf LATTICE2008} (2008) 014
  [arXiv:0811.4146 [hep-lat]], and references therein.

\bibitem{Flynn:2008tg}
  J.~M.~Flynn and C.~T.~Sachrajda  [RBC Collaboration and UKQCD
                  Collaboration],
  Nucl.\ Phys.\  B {\bf 812} (2009) 64
  [arXiv:0809.1229 [hep-ph]].

\bibitem{Bijnens:2009yr}
  J.~Bijnens and A.~Celis,
  Phys.\ Lett.\  B {\bf 680} (2009) 466
  [arXiv:0906.0302 [hep-ph]].

\bibitem{Bijnens:2010ws}
  J.~Bijnens and I.~Jemos,
  Nucl.\ Phys.\  B {\bf 840} (2010) 54
  [arXiv:1006.1197 [hep-ph]],
  Erratum  Nucl.\ Phys.\  B {\bf 844} (2011) 182

\bibitem{Bijnens:1998fm}
  J.~Bijnens, G.~Colangelo and P.~Talavera,
  JHEP {\bf 9805} (1998) 014
  [arXiv:hep-ph/9805389].

\bibitem{Bijnens:1999sh}
  J.~Bijnens, G.~Colangelo and G.~Ecker,
  JHEP {\bf 9902} (1999) 020
  [arXiv:hep-ph/9902437].

\bibitem{Pich}
A.~Pich, Lectures at Les Houches Summer School in
Theoretical Physics, Session 68: Probing the Standard Model of Particle
Interactions, Les Houches, France, 28 Jul - 5 Sep 1997,
[hep-ph/9806303].

\bibitem{Scherer1}
S.~Scherer,
{\em Adv. Nucl. Phys.}, 27 (2002) 277 [hep-ph/0210398].

\bibitem{Wise:1992hn}
  M.~B.~Wise,
  Phys.\ Rev.\  D {\bf 45} (1992) 2188.

\bibitem{Burdman:1992gh}
  G.~Burdman and J.~F.~Donoghue,
  Phys.\ Lett.\  B {\bf 280} (1992) 287.

\bibitem{Goity:1992tp}
  J.~L.~Goity,
  Phys.\ Rev.\  D {\bf 46} (1992) 3929
  [arXiv:hep-ph/9206230].

\bibitem{Wise:1993wa}
  M.~B.~Wise,
  Lectures given at CCAST Symp. on Particle Physics at the Fermi scale,
  arXiv:hep-ph/9306277.

\bibitem{Heavyquarkbook}
  A.~V.~Manohar and M.~B.~Wise,
  Camb.\ Monogr.\ Part.\ Phys.\ Nucl.\ Phys.\ Cosmol.\  {\bf 10} (2000) 1.

\bibitem{SCET}
  S.~Fleming,
  PoS(EFT09)002 [ arXiv:0907.3897 [hep-ph]], and references therein.

\bibitem{GL3}
  J.~Gasser and H.~Leutwyler,
  Nucl.\ Phys.\  B {\bf 250} (1985) 517.

\bibitem{BC}
  J.~Bijnens, F.~Cornet,
  Nucl.\ Phys.\  {\bf B296 } (1988)  557.

\bibitem{Falk:1993fr}
  A.~F.~Falk and B.~Grinstein,
  Nucl.\ Phys.\  B {\bf 416} (1994) 771
  [arXiv:hep-ph/9306310].

\bibitem{Becirevic:2003ad}
  D.~Becirevic, S.~Prelovsek and J.~Zupan,
  Phys.\ Rev.\  D {\bf 68} (2003) 074003
  [arXiv:hep-lat/0305001].

\bibitem{Nakamura:2010zzi}
  KNakamura {\it et al.} [ Particle Data Group Collaboration ],
  J.\ Phys.\ G {\bf G37 } (2010)  075021.

\bibitem{deDivitiis:1998kj}
  G.~M.~de Divitiis, L.~Del Debbio, M.~Di Pierro, J.~M.~Flynn, C.~Michael and J.~Peisa
                  [UKQCD Collaboration],
  JHEP {\bf 9810} (1998) 010
  [arXiv:hep-lat/9807032].

\bibitem{Isgur:1989ed}
  N.~Isgur, M.~B.~Wise,
  Phys.\ Lett.\  {\bf B237 } (1990)  527.

\bibitem{Randall:1993qg}
  L.~Randall, M.~B.~Wise,
  Phys.\ Lett.\  {\bf B303 } (1993)  135-139.
  [hep-ph/9212315].

\bibitem{Politzer:1990ps}
  H.~D.~Politzer,
  Phys.\ Lett.\  B {\bf 250} (1990) 128.

\bibitem{Jenkins:1992qv}
  E.~E.~Jenkins, M.~J.~Savage,
  Phys.\ Lett.\  {\bf B281 } (1992)  331-335.

\bibitem{Boyd:1995pq}
  C.~G.~Boyd, B.~Grinstein,
  Nucl.\ Phys.\  {\bf B451 } (1995)  177-193.
  [hep-ph/9502311].

\bibitem{Eeg:2007ha}
  J.~O.~Eeg, S.~Fajfer, J.~F.~Kamenik,
  JHEP {\bf 0707 } (2007)  078.
  [arXiv:0705.4567 [hep-ph]].

\bibitem{Vermaseren:2000nd}
  J.~A.~M.~Vermaseren,
  arXiv:math-ph/0010025.

\bibitem{Bijnens:2002hp}
  J.~Bijnens and P.~Talavera,
  JHEP {\bf 0203} (2002) 046
  [arXiv:hep-ph/0203049].

\bibitem{Passarino:1978jh}
  G.~Passarino and M.~J.~G.~Veltman,
  Nucl.\ Phys.\  B {\bf 160} (1979) 151.

\bibitem{'tHooft:1978xw}
  G.~'t Hooft and M.~J.~G.~Veltman,
  Nucl.\ Phys.\  B {\bf 153} (1979) 365.
\end{thebibliography}
\end{document}